\documentclass[aps,superscriptaddress,floatfix,onecolumn,preprint]{revtex4-1}

\usepackage{graphicx}
\usepackage{dcolumn}
\usepackage{bm}
\usepackage{amssymb}
\usepackage{psfrag}
\usepackage{subfigure}
\usepackage{color}
\usepackage{amsmath}

\newcommand{\be}{\begin{equation}}
\newcommand{\ee}{\end{equation}}
\newcommand{\bea}{\begin{eqnarray}}
\newcommand{\eea}{\end{eqnarray}}

\newcommand{\bef}{\begin{figure}}
\newcommand{\enf}{\end{figure}}
\newcommand{\p}{\mathbf{p}}
\renewcommand{\r}{\mathbf{r}}

\newcommand{\F}{\mathbf{F}}

\newcommand{\A}{\mathbf{A}}
\newcommand{\Ree}{\mathrm{Re}}
\newcommand{\Imm}{\mathrm{Im}}

\newcommand{\w}{\omega}

\graphicspath{new_figures2/}

\begin{document}
\title{Interpreting Attoclock Measurements of Tunnelling Times }

\author{Lisa Torlina$^{*}$}
\affiliation{Max-Born-Institut, Max-Born-Str. 2A, 12489 Berlin}
\author{Felipe Morales$^{*}$}
\affiliation{Max-Born-Institut, Max-Born-Str. 2A, 12489 Berlin}
\author{Jivesh Kaushal}
\affiliation{Max-Born-Institut, Max-Born-Str. 2A, 12489 Berlin}
\author{Igor Ivanov}
\affiliation{Research School of Physical Sciences, The Australian National University, Canberra ACT 0200, Australia}
\author{Anatoli Kheifets}
\affiliation{Research School of Physical Sciences, The Australian National University, Canberra ACT 0200, Australia}
\author{Alejandro Zielinski}
\affiliation{Ludwig Maximilians University, Theresienstrasse 37, D-80333 Munich, Germany}
\author{Armin Scrinzi}
\affiliation{Ludwig Maximilians University, Theresienstrasse 37, D-80333 Munich, Germany}
\author{Harm Geert Muller}
\affiliation{Max-Born-Institut, Max-Born-Str. 2A, 12489 Berlin}
\author{Suren Sukiasyan}
\affiliation{Department of Physics, Imperial College London, South Kensington Campus, SW7 2AZ London, United Kingdom}
\author{Misha Ivanov}
\affiliation{Max-Born-Institut, Max-Born-Str. 2A, 12489 Berlin}
\affiliation{Department of Physics, Imperial College London, South Kensington Campus, SW7 2AZ London, United Kingdom}
\author{Olga Smirnova}
\affiliation{Max-Born-Institut, Max-Born-Str. 2A, 12489 Berlin}

% \begin{affiliations}
%   \item Max-Born-Institut, Max-Born-Str. 2A, 12489 Berlin
%   \item Research School of Physical Sciences, The Australian National University, Canberra ACT 0200, Australia
%   \item Ludwig Maximilians University, Theresienstrasse 37, D-80333 Munich, Germany
%   \item Department of Physics, Imperial College London, South Kensington Campus, SW7 2AZ London, United Kingdom
% \end{affiliations}

%TC:ignore
\begin{abstract}
Resolving in time the dynamics of light absorption by atoms and molecules, and the electronic rearrangement this induces, is among the most
challenging goals of attosecond spectroscopy. The attoclock is an elegant approach to this problem, which encodes ionization times 
in the strong-field regime. However, the accurate reconstruction of these times from experimental data presents a 
formidable theoretical challenge. Here, we solve this problem by combining  analytical theory with ab-initio numerical simulations. 
We apply our theory to numerical attoclock experiments on the hydrogen atom to extract ionization time 
delays and analyse their nature. Strong field ionization is often viewed as optical tunnelling through the barrier created by the field and 
the core potential. We show that, in the hydrogen atom, optical tunnelling is instantaneous. 
By calibrating the attoclock using the hydrogen atom, our method opens the way to identify 
{possible delays associated with multielectron dynamics 
during strong-field ionization.}

\end{abstract}
\maketitle
%TC:endignore

\section{Introduction}

Advances in attosecond technology have opened up the intriguing opportunity to time electron release during
photoionization.  New experimental techniques such as the attosecond streak camera\cite{Schultze}, 
high harmonic spectroscopy\cite{nirit}, attosecond transient absorption\cite{Goulielmakis} and the 
attoclock\cite{keller1,keller2,keller3,keller4} are now able to provide the exceptional time-resolution -- down to the level
%of about ten attoseconds, which is needed to time-resolve ionization.  The removal of an electron from an atom
%ASK
of tens of attoseconds 
(1~asec = 10$^{-18}$~s) 
-- needed to time-resolve ionization.  The removal of an electron from an atom
or molecule during one-photon ionization creates a non-equilibrium charge distribution which evolves 
on the attosecond time scale\cite{Cederbaum}. Ionization time then serves as a sensitive measure encoding the dynamics of core
rearrangement triggered by electron removal (see e.g.\cite{kheifets2,taylor}).

While the use of intense  IR fields  as either pump or probe in time-resolved ionization experiments 
provides access to the  time scale of electronic motion,  it also introduces a hurdle  in interpreting  
such experiments\cite{klunder,mivanov,joachim,kheifets2,taylor,Maquet}.
Identifying and disentangling time delays related to multielectron dynamics from
the apparent delays induced by the interaction with the
IR field  is  challenging both technically and conceptually.
In one-photon ionization \cite{Schultze}, understanding the nature of the measured delays required the 
accurate calibration of the measurement schemes, with the hydrogen atom used as a benchmark, see e.g. \cite{klunder,mivanov,joachim,Maquet}.

Looking beyond the weak field one-photon case, multiphoton ionization can also excite rich multielectron 
dynamics, which calls for the accurate measurement of ionization times in the strong field regime.  
What's more, strong field ionization is often viewed as a tunnelling process, where the bound electron 
passes through the barrier created by the laser field and the core potential. Consequently, time resolving 
this process opens the intriguing opportunity \cite{keller1,keller2,keller4,nirit} to revisit the long-standing problem of tunnelling times.

The measurement of tunnelling times in strong-field ionization has been pioneered by the group of 
U. Keller  \cite{keller1,keller2,keller3,keller4} using the attoclock technique. The attoclock set-up measures 
angle- and energy- resolved photoelectron spectra produced by ionization in strong, 
nearly circularly polarized infrared  (IR) fields. Essentially, the rotating electric field vector serves as 
the hand of a clock, deflecting electrons in different directions depending on their moment of escape from the atom. 
The tunnelling perspective provides a simple picture of how this works. The strong circularly polarized 
field combined with the binding potential of the atom together create a rotating barrier through which an 
electron can tunnel (Fig.1a). Due to the rotation of the barrier, the electron tunnels in
different directions at different times, and is subsequently detected at different angles after the end of the pulse (Fig.1b).

\begin{figure}
\includegraphics[width=5in,angle=0]{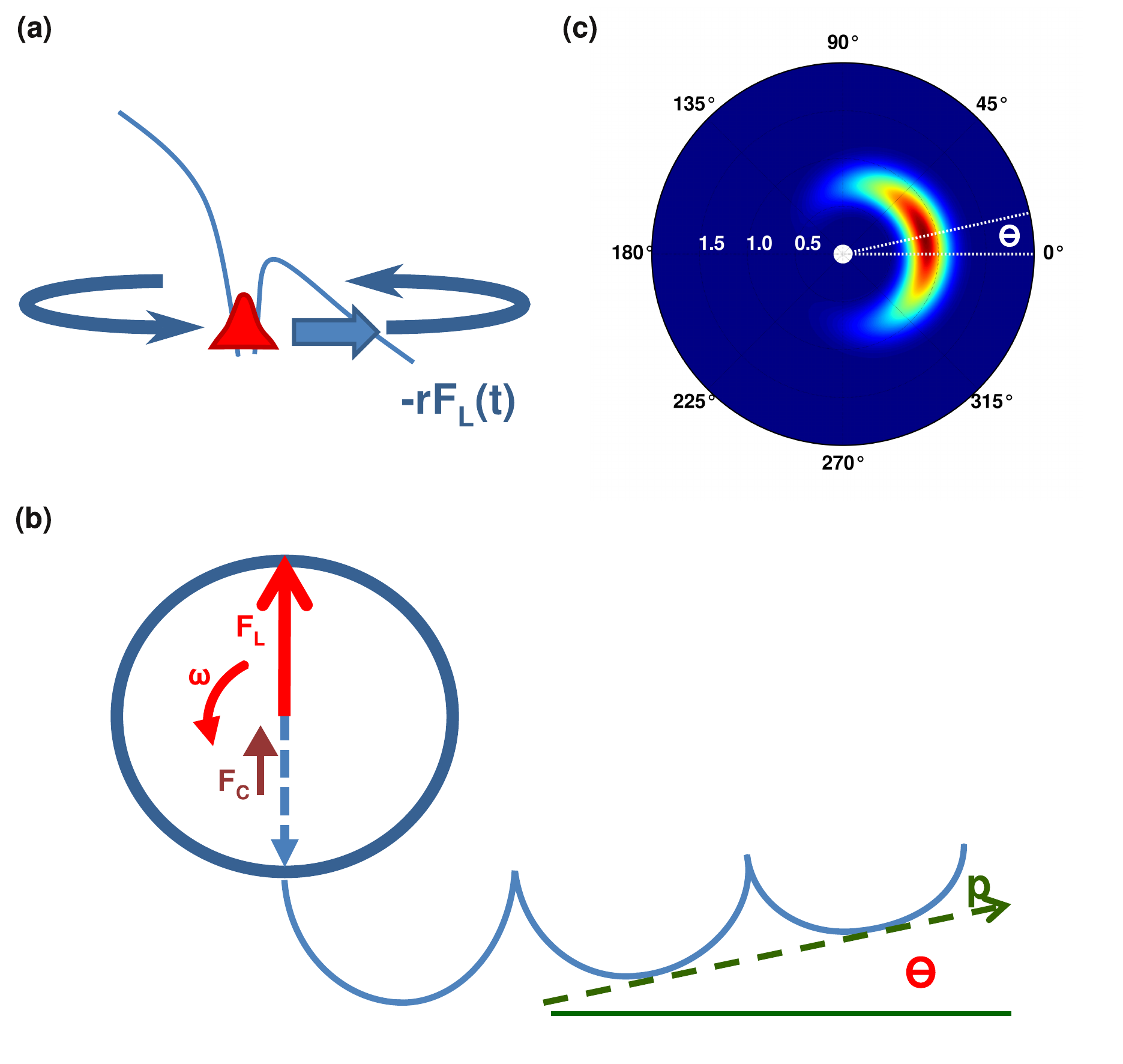}
\caption{The attoclock setup. (a) The tunnelling perspective on the attoclock: 
the laser field and the core potential together create a rotating barrier through which the 
bound electron can tunnel. As the barrier rotates, the electron will escape in different directions at different times. 
(b) A cartoon illustrating our ionization geometry. The laser field rotates counter-clockwise and 
reaches its maximum value when the electric field $\mathbf{F}_L$ points up at an angle of 90$^\circ$.
When the electric field points up, the electron tunnels down, and in the absence of electron-core interaction, 
we would expect to detect it at an angle of $\phi=0^\circ$. 
An offset $\theta$ from this angle could be due to the attractive potential of the core (force $\mathbf{F}_C$) as 
shown in the figure, and/or possible tunnelling delays.  
  (c) The experimental observable: the angle- and energy- resolved photoelectron spectrum, 
shown for ionization from the ground state of the hydrogen atom induced by a 
single-cycle circularly polarized infrared pulse. Dashed lines show the offset angle $\theta$.
}  \label{Fig1}
\end{figure}

Suppose the field rotates counterclockwise and reaches its maximum at $t=0$ 
when the field vector $\F_L(t)$ points at an angle of $\phi=90^\circ$ (Fig.1b). 
In the tunnelling picture, this instant is associated with the thinnest tunnelling barrier and the highest probability of 
ionization. In the absence of (i) electron-core interaction after tunnelling and (ii) tunnelling delays, we 
expect that an electron which escapes at time $t_0=0$ will be detected at an angle of $\phi=0^\circ$, orthogonal to
$\F_L(t)$. Indeed, if the electron is released from the barrier with zero initial velocity as suggested by the tunneling picture 
(Fig.1(a)), its final momentum at the detector will be $\p=-\A_L(t_i)$, where $\A_L(t_i)$ is the vector-potential 
of the laser field at the moment of ionization. For circularly polarized pulses, $\A_L$ is orthogonal to $\F_L$ (up to effects of the ultrashort envelope). 

An observed deviation of the photoelectron distribution maximum from $\phi=0^\circ$ could 
come from the deflection of the outgoing electron by the attractive core potential (Fig.1b) and, 
possibly, from tunnelling delays\cite{keller1,keller2}. This deviation is characterised by the offset angle $\theta$ (Fig.1c). 
Experimentally, $\theta$ can be measured with high accuracy ($\delta \theta\sim 2^\circ$), which implies 
the potential to measure ionization delays with  accuracy  $\delta \tau=\delta \theta/\omega_L\sim 15$~asec for 800~nm radiation.

{However, the reconstruction of ionization times from experimentally measured offset angles is  sensitive to the assumptions made about the underlying process. To date, the theoretical approaches used to interpret attoclock results have relied on three assumptions}
{\cite{keller1,keller2,keller3,keller4,kheifets}. 
(A1) First, based on exponential sensitivity of strong-field ionization to the electric field, 
it is assumed that the highest probability for the electron to tunnel is at the peak of the electric field. 
(A2) Second, ionization is assumed to be completed once the electron emerges from the barrier. 
(A3) Third, electron dynamics after the barrier exit are described classically, 
assuming some point of exit and initial distribution of velocities \cite{keller4,prl_keller2}.}
%Typically, the following assumptions are made:
% \begin{itemize}
% \item A1: The exponential sensitivity of strong-field ionization to the electric
%field implies that there is a preferred time and direction of ionization. It is postulated that the highest probability for the electron to tunnel is at the peak of the electric field.
%\item  A2: Ionization is assumed to be completed once the electron emerges from the barrier.
%\item A3: Electron dynamics after the barrier exit are described classically, assuming some point of exit and initial distribution of velocities \cite{keller4,prl_keller2}.
% \end{itemize}
Within this classical model, the accuracy of extracting time delays from attoclock 
measurements  depends on the initial conditions assumed for the classical electron  dynamics. 
These initial conditions, however, cannot be established unambiguously.
The resulting ambiguity in interpreting attoclock measurements is a major 
bottleneck for reconstructing ionization times with attosecond precision. 

In light of this, we provide a consistent interpretation and calibration for attoclock measurements 
of ionization times, making no ad hoc assumptions. We do this by  combining  analytical theory with ab-initio simulations. 
To calibrate the attoclock, we focus on the hydrogen atom. Doing so, we (i) find very good agreement between our analytical theory and numerical experiments, 
(ii) show that, for one-electron systems, purely tunnelling delays during strong-field 
ionization are equal to zero and 
(iii) reconstruct ionization times for the hydrogen atom, finding 
deviations from the conventional tunneling picture expressed by assumptions A1-A3.
{We also show how the calibration based on single active electron 
dynamics  can be used to identify multielectron contributions 
to the attoclock observable in multi-electron systems.}

 \section{Results}

\subsection{Theoretical description.}

{Our theoretical approach is based on the Analytical R-Matrix (ARM) method \cite{arm1,arm2,jivesh,lisa}. The key mathematical approximations of this theory and its application to nearly single-cycle pulses are described in the Appendix.
}
%A detailed description of the application of ARM to nearly single-cycle pulses,
%including expressions for angle- and energy-resolved electron spectra, 
%is given in the Supplementary Material (SM). 
%Crucially, ARM allows us to establish a 
%connection between the attoclock observables
%and the ionization times.  
{In ARM, the probability  $w(p,\phi)$
of detecting an
electron at an angle $\phi$ with momentum $p$
is described by an integral over all possible instants of ionization\cite{arm1,arm2,jivesh,lisa}.
By calculating this time-integral using the saddle point method, we express $w(p,\phi)$ 
via the contribution of  'quantum trajectories' that start from the 
atom at complex times $t_s(\phi,p)$.
Mathematically, 
%application of the saddle-point method  
%in the present context is equivalent to 
%the Wentzel-Kramers-Brillouin (WKB) approach 
%in the time-domain. 
the saddle point approach is 
justified when the electron action $S$ accumulated
along the 'quantum trajectory' is large, $S\gg \hbar$,
which is naturally satisfied in strong low-frequency fields.
ARM also requires sufficiently thick tunnelling barriers,
restricting  analytical work to circularly polarized fields with 
intensities $I<4\times 10^{14}$W/cm$^2$.
Comparison with ab-initio calculations allows us to judge the accuracy of
the analytical approach.}

{Since time arises naturally along each 'quantum trajectory', 
establishing a connection between the attoclock observable $w(\phi,p)$ and the 
associated time  $t_s(\phi,p)$ becomes possible. }
It is this mathematically established connection that allows us to 
calibrate the attoclock. 
The real part $t_i(\phi,p)\equiv {\rm Re}t_s(\phi,p)$
is  the 'ionization time'. In the 
tunnelling picture, this time corresponds to the moment 
at which the electron emerges in the classically allowed region. For a 'zero-range' binding potential, which 
supports only a single bound state,
we obtain:
\begin{equation}
	{\rm Re}t^{(0)}_s(\phi,p)\equiv t^{(0)}_i(\phi,p) 
= \frac{\phi}{\omega}+\Delta t_i^\mathrm{env}(\phi,p).
\label{eq:tiShortRange}
\end{equation}
Here the superscript $(0)$ denotes the `zero-range' potential.
The small correction $\Delta t_i^\mathrm{env}$ is due to 
the ultrashort pulse envelope. 
It accounts for the fact that, for very short circular pulses, 
the electric field and the vector potential are not orthogonal at all times during the pulse; 
 $\Delta t_i^\mathrm{env}$ disappears for ionization at the peak of the field.  
In the geometry of Fig.1, the offset angle $\theta$ is the angle $\phi$ associated 
with the highest photoelectron signal. 

For an arbitrary potential, ARM yields (see Appendix):
\begin{equation}
	 t_i(\phi,p) = \frac{\phi}{\omega} + \Delta t_i^\mathrm{env}(\phi,p)
+ \Delta t_i^C(\phi,p),
\label{eq:ti}
\end{equation}
where $\Delta t_i^C= \Ree \Delta t_s^C$ is given by the following expression:
\begin{equation}
        \Delta t_s^C=-\frac{d W_C(\phi,p)}{d I_p},
	\label{eq:Delay} 
\end{equation}
Here $W_C$ is the phase acquired by the laser-driven electron due to its 
interaction with the core and $I_p$ is the ionization potential (see Appendix for detailed derivation and 
explicit expression for $W_C$, including its dependence on the core potential). 
%Eq.(\ref{eq:Delay})  gives the additional 
%ionization delay compared to the `zero-range' potential, i.e. compared 
%to the propagation of the electron with the same final %momentum in the laser field and in the free space:
%is our key result: it is the analogue of the Wigner-Smith-type photoionization delay for one-photon ionization, obtained here for %the strong-field regime.  
%Similar to the one-photon case, it is given by the derivative of the electron photo-ionization phase.
%In the one-photon case, the  delay is given by the derivative of the phase of the photoionization matrix element 
%with respect to the photoelectron energy. In contrast to scattering problems, this phase is finite and converges even for the long range potential 
%due to the finite extent of the bound state.
The derivative is taken with respect to the electron binding energy $I_p$, keeping the initial
and the final electron momenta constant (see Appendix). 
%We note that, in contrast to scattering problems,  the phase required to obtain ionization delay 
%is finite in both one-photon and %strong field cases (see SM). 
%%%%%%%%%%%%%%%END OF WHY DO WE NEED ARM%%%%%%%%%%%%%%%%%%%%%%%%%%%

%EXPLAIN WHY DO WE DO IT: MERITS OF numerical PROCEDURE vs experiment!
\subsection{Attoclock ab-initio.}
We can now (i) test the ability of ARM to quantitatively describe 
attoclock measurements, (ii) apply the results Eq. (\ref{eq:ti},\ref{eq:Delay})
to reconstruct ionization times $t_i$,
(iii) investigate the presence of tunnelling delays associated with the electron's motion in the classically forbidden
region. To  this end,  we perform ab-initio numerical simulations of attoclock measurements for a benchmark system:
a hydrogen atom interacting with a perfectly circularly polarized, nearly 
single-cycle laser pulse with central wavelength $\lambda=$ 800 nm. 
{The merit of using ab-initio simulations for hydrogen is the 
unprecedented accuracy this affords when analysing the
attoclock observables: the hydrogen atom is unique in allowing an exact numerical 
solution of the full time-dependent Schroedinger equation (TDSE) 
in a circularly polarized field,
requiring no approximations beyond the standard non-relativistic 
and dipole approximations.
Approaching the problem numerically gives us full
control of all pulse parameters -- intensity, ellipticity, pulse shape and carrier-envelope phase -- which is 
important when time-resolving highly nonlinear processes at the 10 asec
level.}

Since every numerical scheme must deal with convergence issues related to the 
finite discretization step, the size of the simulation box, time-propagation routines, etc,
we compare three independent calculations of the angle- and energy- resolved photoelectron spectra, 
done using three different methods and propagation algorithms \cite{muller,tao12,kheifets} (see Methods). 
The results are in very good agreement.
We then compare the numerical results with the analytical theory, 
to check its validity, and again find 
very good agreement across a wide range of intensities. Fig.2 shows this comparison for 
$I=$1.75, 2.5 and 3.4$\times10^{14}$ W/cm$^2$.  
The laser field is defined by $\F_L(t)=-{\partial\mathbf{A}_L(t)}/{\partial t}$, where
\begin{eqnarray}
      \mathbf{A}_L(t) = -A_0 \ \cos^4(\w t/4) \ (\cos(\w t) \ \mathbf{\hat{x}} + \sin(\w t) \ \mathbf{\hat{y}}).
 \label{pulse}
\end{eqnarray}
The field rotates counter-clockwise and points at an angle of $90^{\circ}$ when it reaches its maximum at time $t=0$. 

The offset angle $\theta$ is extracted by finding the peak of the photoelectron distribution. 
Fig.3a shows the offset angles calculated using the three numerical methods and the ARM approach as a function of laser intensity. The numerical
results agree within $0.5^\circ$, and the deviation between the analytical and numerical results is within 
$2^\circ$. This slight discrepancy is analysed further in Fig.3(b), where we zoom into the region of intensities between 
$1\times 10^{14}$W/cm$^2$ and $3\times 10^{14}$W/cm$^2$. 
The vertical lines indicate the angles at which the distribution falls by a mere 0.1\% compared to the peak of the signal intensity.
Within this deviation, the analytical and numerical offset angles agree.
These vertical lines highlight the extremely flat nature of the distribution around the maximum, even for the single-cycle pulse we have used and gauge the accuracy one has to reach to locate the maximum of photoelectron distribution.
The flatness of the spectrum we see here may also challenge the accuracy of identifying $\theta$ in experiments.

\begin{figure}
\includegraphics[width=5in,angle=0]{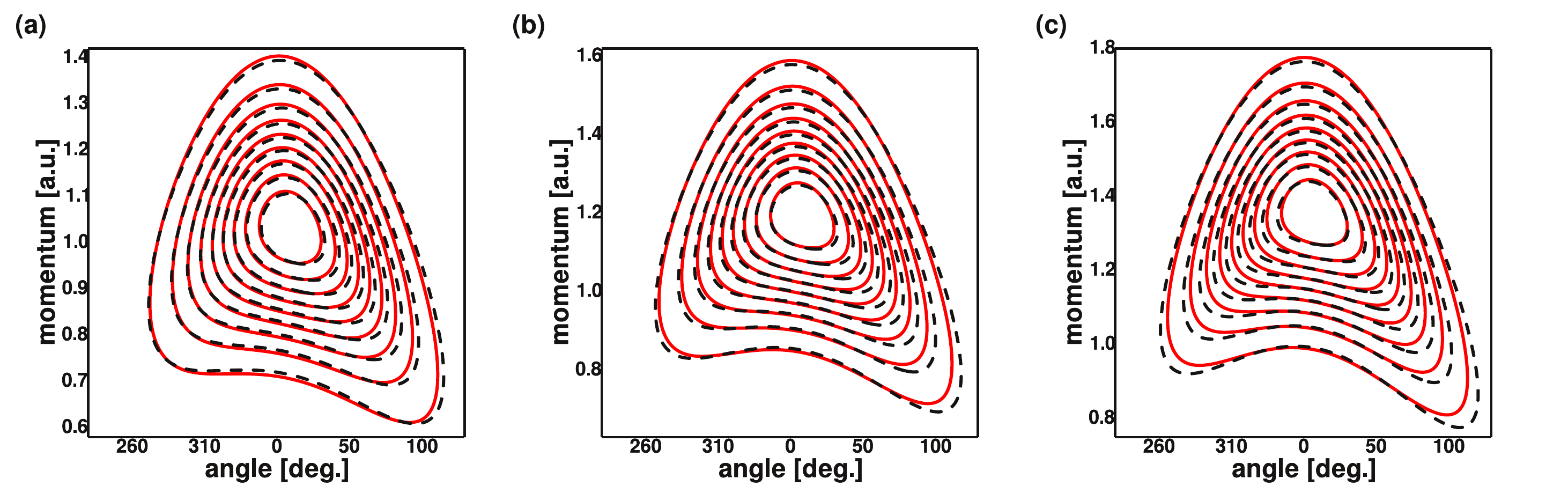}

\caption{Angle- and energy- resolved photoelectron spectra produced by the 
strong field ionization of the hydrogen atom using a single-cycle circularly polarized laser pulse with 
wavelength $\lambda=800$ nm and intensity (a) $I=1.75\times 10^{14}$ W/cm$^2$, 
(b) $I = 2.5\times 10^{14}$ W/cm$^2$, and (c) $I=3.4 \times 10^{14}$ W/cm$^2$. 
The form of the laser pulse is specified in Eq.\eqref{pulse}. Solid red contours show spectra 
obtained analytically using the ARM theory. Dashed black contours are the results of ab initio numerical 
calculations performed using the method labelled TDSE H1 (see Methods section). 
The distributions are normalized to 1, and contours correspond to signal intensity changing 
from 0.1 to 0.9 in steps of 0.1, with the innermost contour at 0.9.
} \label{Fig2}
\end{figure}

\begin{figure}
\includegraphics[width=5in,angle=0]{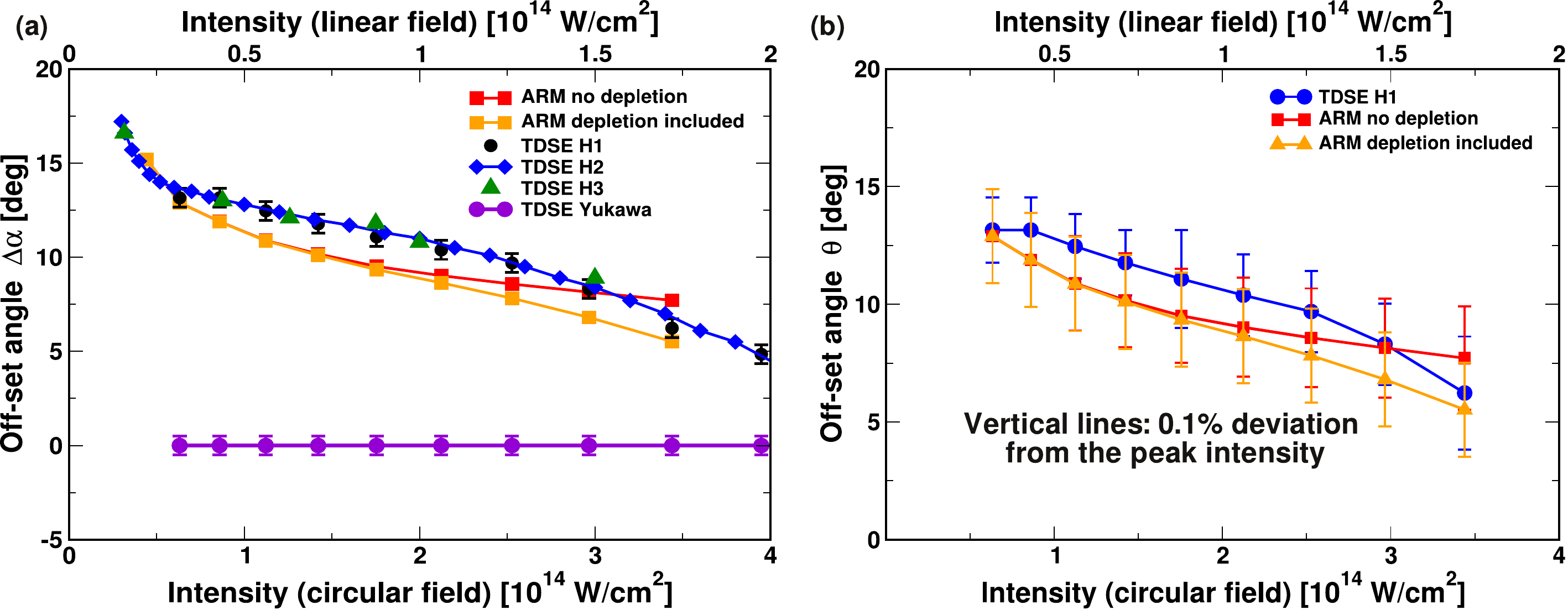}

\caption{Offset angles $\theta$ extracted from
photelectron spectra as a function of intensity.
(a) A comparison of the offset angles obtained for the hydrogen atom using the three different numerical 
methods (black circles, blue diamonds and green triangles correspond to TDSE H1, H2 and H3 respectively, see Methods) 
and the ARM theory (red and yellow squares, the latter include effect of the ground state depletion,
see Appendix for details). Violet circles show the numerically obtained offset angles for the short-range Yukawa potential.
(b) A close up of the analytical (red and yellow squares) and numerical (blue circles, TDSE H 1) results
for the offset angle for hydrogen. The vertical lines indicate the angles at which the signal intensity 
is reduced by a mere 0.1\% compared to the peak value.
%(c) Reading the ionization time from the attoclock observable $\theta$. 
%The attoclock observable (black circles, TDSE H 1 and TDSE H 2) provides access to the ionization time 
%$t_i=\Ree  t_s$ (green triangles) by subtracting the ionization delay $\Delta t_i$ (blue inverted triangles) calculated 
%for $\p_{max}$ obtained numerically from the ab-initio results corrected for the effects of the ultra-short 
%envelope (orange diamonds). %The error bars $\pm 1.5 ^{o}$ on green curve reflect the accuracy of analytical 
%modelling of $\theta$, evaluated by comparison with numerical results (Fig.3 (b)).
%(d) Cartoon illustrating attoclock measurements and the physical
%meaning of real part of the saddle point solution.
%Attoclock measures full ionization time, shown by red arrow. Ionization starts at time $t_i=\Ree t_s$ (green arrow).
%Ionization occurs after field maximum (orange arrow). Ionization delay is shown by blue triangle.
} \label{Fig3}
\end{figure}

All calculations show a very interesting trend in intensity.  At lower intensities, when the barrier for tunnelling is thicker, there is a bigger deflection angle.
Does this trend represent a tunnelling delay, as suggested recently \cite{prl_keller2, landsman}?

\subsection{Delays in tunnelling.}
In the hydrogen atom, the angular offset may come from two sources: tunnelling delay and the interaction 
between the departing electron and the nucleus. 
{As a first step towards distinguishing these two possibilities, we} replace the Coulomb potential of the hydrogen atom by a short range potential.
%One way in which we can distinguish these two possibilities is 
%by replacing the Coulomb potential of the hydrogen atom by a short range potential.  
In this case, 
the tunnelling barrier will still be present; however, electron-nucleus interaction after tunnelling is  turned off. 
To investigate this, the numerical calculations were repeated for a short-range Yukawa potential, 
$U_{Y}=-Ze^{-r/a}/r$, with $Z=1.94$ and $a=1.0$ a.u. chosen to yield the ionization potential of the hydrogen atom. 
The results are summarized in Fig.3a and Fig.4. At all intensities, we find that the offset angle $\theta$ is equal to zero in this instance. 
That is, the attoclock measures no tunnelling delays for the short-range Yukawa potential. 
{We now move to the hydrogen atom, where the presence of multiple
excited states can, in principle, alter
the tunneling process via electronic excitations before tunnelling.}
%Therefore,  for the hydrogen atom, the offset 
%angle must be determined by the interaction of the outgoing 
%laser-driven electron with the core potential, and not by tunnelling delays.

\begin{figure}
\includegraphics[width=5in,angle=0]{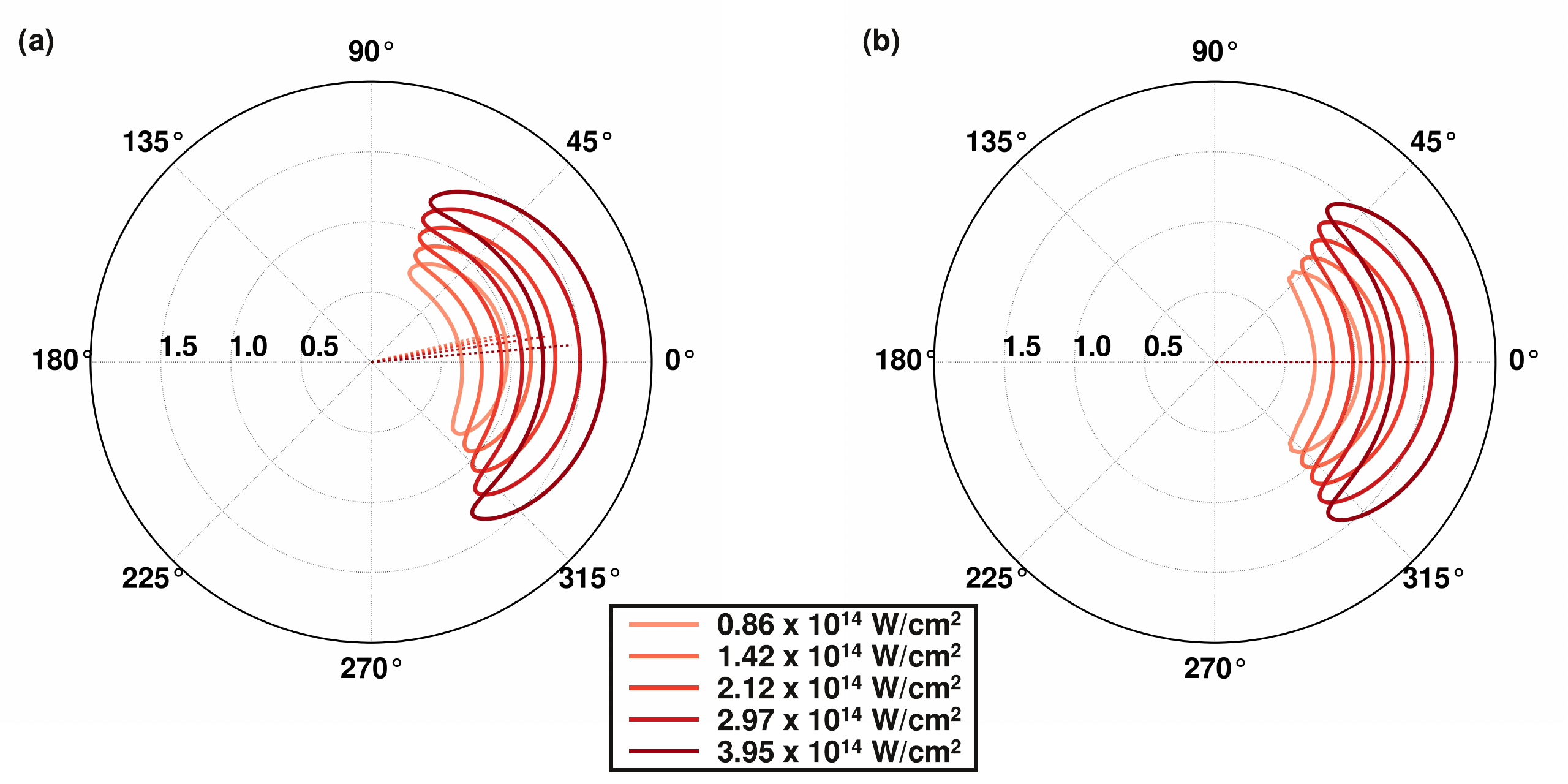}
\caption{A comparison of the photoelectron spectra calculated numerically using method TDSE H1 for (a) 
hydrogen and (b) the short-range Yukawa potential. Each contour corresponds to the same signal strength, 
but a different laser intensity. Pulse shape and wavelength are the same as in Fig.2 and 3. In (b), 
the offset angle is zero at all intensities, and hence the attoclock measures no tunneling delays for the Yukawa potential.
%Circularly polarized single cycle laser pulse is specified in Eqs.(\ref{Fig1},\ref{envelope}).
%Panel (a)
%(a) I=1.75 10$^{14}$ W/cm$^2$, (b) I=2.5 10$^{14}$ W/cm$^2$, (c) I=3.4 10$^{14}$ W/cm$^2$. (d) Off-set angle vs intensity
} \label{Fig4}
\end{figure}

%We can now apply our analytical theory to investigate the underlying physics. 
%Our discussion in section \ref{analyticalapproach} led to an equation 
%connecting the electron's momentum and angle of detection to ionization time, based on the ARM theory (Eq.\eqref{eq:ti}). 

\subsection{Reconstruction of ionization times in hydrogen.}
Having demonstrated very good agreement between photoelectron spectra calculated using the ARM method and ab initio 
TDSE calculations, we can now  apply the mapping (Eq.\eqref{eq:ti}) to reconstruct ionization times from ab-initio data. 
In particular, for a given photoelectron spectrum, we extract the most probable time of ionization by evaluating 
Eq.\eqref{eq:ti} at the spectrum peak $(\theta,p_\mathrm{peak})$:
\begin{equation}
	t_i(\theta,p_\mathrm{peak}) = \frac{\theta}{\omega} - |\Delta t_i^\mathrm{env}(\theta,p_\mathrm{peak})| - |\Delta t_i^C(\theta,p_\mathrm{peak})|,
	\label{eq:reconstruction}
\end{equation}
where we have used the fact that $\Delta t_i^C$ is negative and 
$\Delta t_i^\mathrm{env}<0$ for $\theta>0$.

\begin{figure}
\includegraphics[width=5in,angle=0]{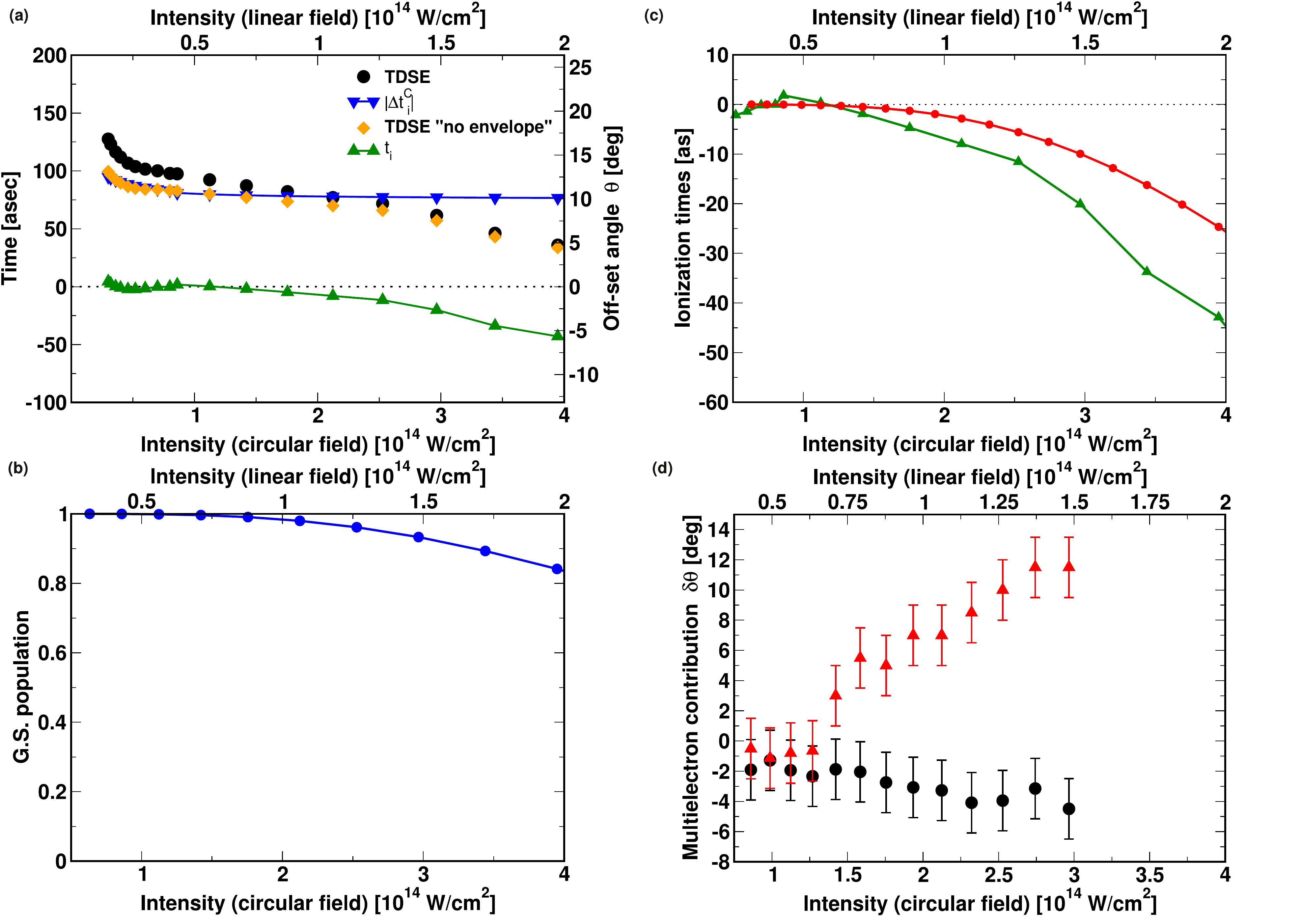}
\caption{Reconstruction of ionization times. (a) Ionization times (left axis) reconstructed using the 
ARM theory from offset angles (right axis) obtained numerically using methods TDSE H1 and H2. 
Black circles are the numerically calculated offset angles, divided by the laser frequency, $\theta/\omega$. 
Orange diamonds show the offset angles with the correction due to the pulse envelope subtracted, 
$t_i^0 = \theta/\omega - |\Delta t_i^\mathrm{env}(\theta,p_\mathrm{peak})|$. 
Blue inverted triangles show the Coulomb correction to the ionization time evaluated at the peak 
of the photoelectron distribution, $|\Delta t_i^C(\theta,p_\mathrm{peak})|$. Green triangles 
show the ionization times we obtain by applying reconstruction procedure defined by Eq.\eqref{eq:reconstruction}. 
In terms of the figure, this is simply the result of subtracting the blue curve from the the orange curve. 
(b) Population of the ground state of the H-atom after the end of the laser pulse as a function of intensity. (c) 
Ionization times reconstructed using the ARM theory, based on numerically derived offset angles 
(green triangles), vs corrections to zero ionization times due to effects of depletion alone (red circles) (see Appendix for details).
(d) Identification of multi-electron effects in attoclock measurements: difference $\delta\theta=\theta_2-\theta_1$ 
between the deflection angles for the two-electron ($\theta_2$) and 
the one-electron ($\theta_1$) systems with the same ionization potential. 
Black dots: results for a 'rigid' model system with high energy of 
ionic excitations $\Delta E=0.47$a.u. Red triangles: results for a model system
with reduced energy of 
ionic excitations $\Delta E=0.21$a.u.} \label{Fig5}
\end{figure}

Fig.5(a) shows the results of applying this reconstruction procedure for offset angles $\theta$ and 
momenta $p_\mathrm{peak}$ obtained numerically using methods TDSE H1 and H2 (see Methods). 
Black circles represent the first term in Eq.\eqref{eq:reconstruction} above: the numerically calculated 
offset angles, divided by the laser frequency. They correspond to the ionization times we would have reconstructed 
had we neglected the Coulomb effects and assumed the long pulse limit: $t_i^0=\theta/\omega$. 
Orange diamonds represent the above offset angles with the envelope correction 
$|\Delta t_i^\mathrm{env}|$ subtracted: 
the first two terms in Eq.\eqref{eq:reconstruction}. 
Essentially, the envelope correction removes the effects of pulse shape from the data: 
within the analytical approach, we have verified that offset angles corrected in this way 
become virtually independent of the shape of the envelope used. 
Blue inverted triangles show the Coulomb correction to the ionization time, the final term in 
Eq.\eqref{eq:reconstruction}. 
%As explained above, 
%this is the analogue of the Wigner-Smith `ionization delay' for the 
%strong-field ionization regime. 
Finally, green triangles show the reconstructed 
ionization times $t_i$ themselves.

Across all intensities, we find that the reconstructed ionization times are never positive. 
%The electron therefore appears in the continuum either at the
% peak of the field or earlier (for $I > 1.5\times 10^{14}$ 
%W/cm$^2$), but never later. 
The absence of such positive times, in turn, implies the {absence of tunnelling delays 
in the strong field ionization of the hydrogen atom in IR fields. }
%{Note that circular polarization and infrared frequency of the laser field 
% minimize real excitations of  bound states. Existence of such excitations prior 
%to ionization could affect  ionization dynamics.}

For $I > 1.5\times 10^{14}$ W/cm$^2$, ionization times become negative, 
which indicates that the dominant contribution to ionization occurs just before the field reaches its maximum. What could the origin of 
this effect be?
One possible explanation is the depletion of the ground state: a loss of population before the peak 
of the field would enhance the relative contribution of early ionization events, decreasing the off-set angle as shown in Fig. 3(a) within the ARM theory. 
The depletion of the ground state, calculated numerically, is shown in Fig. 5(b) (see Appendix for details). As expected,
depletion increases with intensity, which in turn should give rise to negative $t_i$. 
However, as Fig. 5(c) shows, if we calculate the expected negative shift of the 
ionization times based on depletion alone (red circles) (see Appendix), we find that it is not 
sufficient to explain the observed negative ionization times (green triangles). 
{This implies that 
either the analytical expression Eq.(\ref{eq:Delay})  
 becomes less accurate at higher intensities,
or {there is another} physical effect at play. 
{The latter possibility is explored in the Appendix using additional numerical tests, which are entirely independent of the analytical theory.}}
These tests
confirm the appearance of negative ionization  times and show that they
may be related to 
%trapping of electrons 
%into excited states after tunnelling.
%A similar effect known as 
`frustrated tunnelling', a phenomenon well documented for the case of ionization in  linearly polarized pulses \cite{Ulli}. 
Recent theoretical work has suggested that this can also occur in circular fields 
\cite{uzer1,eberly}, and the negative ionization times we reconstruct may be a signature of this.

In one-photon ionization, the accurate calibration of time delays for the 
hydrogen atom has made it possible to access delays associated with multielectron effects 
\cite{klunder,mivanov,joachim,eva,kheifets3}.
The same applies to multiphoton ionization time-resolved by the attoclock.
It is natural to expect that, as in the one photon case, these may also lead 
to delays during multiphoton ionization in strong laser fields.
The ability to account for the time-shift
$\Delta t_i^C$ asociated with single-electron dynamics  
allows one to identify multielectron contributions in attoclock measurements.
%, which will 
%manifest themselves as deviations from the single-electron ionization times calibrated here.

{Fig. 5 (d) shows the difference $\delta \theta$ between the numerically calculated offset angles ($\theta_2,\theta_1$) for a two-electron and a single electron system with the same binding energy and the same long-range core potential, $\delta \theta=\theta_2-\theta_1$ 
(see Appendix for details). We observe two complementary effects. 
The first is core polarization\cite{keller4}, which
shields the core and therefore slightly {\it reduces} the deflection angle 
caused by electron-core attraction. 
{This leads to 
negative $\delta \theta$ (see Fig. 5(d), black circles).} 
{This is the only effect we see in our numerical simulations when}
electronic excitations in the ion
lie far above the 
ionization threshold ($\Delta E=0.47$ a.u. in our model system).} %In this case, low laser frequency
%reduces the multi-electron response to adiabatic polarization of the core \cite{keller4}.}

{However, if we reduce the energy of electronic excitation in the ion (to $\Delta E=0.21$ a.u), and thereby also the energy of two-electron excitations, while keeping the ionization potential of the neutral fixed at $I_p=0.5$a.u., the picture changes (Fig. 5d, red triangles). As laser intensity increases,}
{
the multielectron correction $\delta \theta$ of the deflection angle starts to rise quickly and becomes positive. This sharp increase of $\delta \theta$ coincides with onset of double ionization.}

{This trend is accompanied by a decrease in the relative delay between the removal of the two electrons during double ionization (see Appendix Fig. 10), and is consistent with
two-electron excitations formed during the first ionization step \cite{arm2,keller3}
(termed 'pre-collision' in \cite{keller3}). Thus, electronic excitations during ionization
may indeed lead to additional positive delays in attoclock measurements of single ionization.}

\section{Conclusions and Outlook}

{As we have seen, because the ARM method naturally includes the concept of trajectories, it makes it possible to reconstruct ionization times from the experimentally observed offset angles. Applying this method to the single electron system, the hydrogen atom, in turn, has allowed us to calibrate the attoclock, revealing the contribution coming purely from the attractive force of the core. We can conclude that any additional offset observed in multielectron systems must then be due to multielectron effects.}

{The presence of trajectories within ARM also makes it possible to
assess the accuracy of the commonly used assumptions (A1)-(A3).}
In contrast to (A3), 
trajectories in ARM are never fully classical. Although the measured quantity (the electron momentum) 
is real, the trajectories
retain an imaginary component of the coordinate all the way to the detector (see Appendix).
This property is directly related to the fact that, for long range potentials, 
ionization is not yet completed at the moment the electron exits the tunnelling barrier, in contrast to (A2), 
see discussion and ab-initio numerical tests in Appendix. 
At high intensities,
the dominant contribution to the photoelectron spectrum may come from ionization that starts 
before the maximum of the electric field, even after effects of depletion are taken into account, 
in contrast to (A1). 
{One should therefore be  cautious when 
using assumptions (A1)-(A3) to interpret attoclock measurements at the $\sim 10$ asec level. }

%To date, the interpretations of attoclock experiments have relied on these assumptions. 
%Our work shows that, at the accuracy of needed to resolve ionization times, these assumptions cannot be used.

%\subsection{Multi-electron effects in tunnelling.}

{
{Our results indicate no tunnelling delays associated with the ionization of the ground-state of the hydrogen atom by a strong low-frequency field. However, the situation may be different when real electronic excitations during ionization are involved. For two-electron systems, our results have shown that the two types of multielectron response to the infrared laser field -- namely the adiabatic polarization of the electronic cloud and real two-electron excitations -- leave distinct and different traces in attoclock measurements, leading to additional delays, either negative or positive.}
Thus, attoclock experiments with molecules or alkaline-earth
atoms, where doubly excited states lie below the first ionization threshold, may uncover rich
multi-electron response manifested in non-trivial, intensity dependent ionization delays
caused by correlation-driven excitations during strong-field ionization.
}

%TC:ignore

\section{Methods}

Our numerical simulations have used three different algorithms to produce ab-initio spectra of strong-field ionization
induced by a nearly single-cycle 800 nm laser pulse. The data labeled 'TDSE H 1' (F. Morales and H. G. Muller) have used the numerical procedure and
the code described in detail in \cite{muller}. The data labeled 'TDSE H 2' (A. Zielinski and A. Scrinzi) were obtained using the t-SURFF
method described in \cite{tao12}. The data labelled 'TDSE H 3' (I. Ivanov and A. Kheifets) have used the numerical procedure and
the code described in detail in \cite{kheifets}.

The method used for the calculations labeled  'TDSE H 1' (F. Morales and H. G. Muller) has been monitored for convergence
by changing the maximum angular momentum up to $L_{max}=120$, while the radial grid size was
increased up to $r_{max}$=2700 a.u.  The spectrum was obtained by projection on the
exact field-free continuum states of the H-atom after the end of the laser pulse. The step size of radial grid was
$\delta r$=0.15 a.u. and the time-step was $\delta t$=0.05 a.u. Convergence was monitored by varying
$\delta r$ down to 0.05 a.u.  and the time-step $\delta t$ down to 0.04 a.u.

T-SURFF ('TDSE H 2', A. Zielinski and A. Scrinzi) combines numerical solutions in an
inner region with approximate analytical solutions in terms of Volkov states
outside. The method is efficient since the numerical part of the solution can be
kept comparatively small: converged results were obtained with an inner region
%$|\vec{r}<r_{\rm max}|=120 a.u.$ using  a finite-element radial discretization with 310 coefficients and
%ASK
$|\vec{r}|<r_{\rm max} =120$~a.u. using  a finite-element radial discretization with 310 coefficients and
an expansion into spherical harmonics up to $L_{\rm max}=95$. The dominant error in
the offset-angle $\theta$ arises from the absence of electron-ion interaction in the Volkov
states. It is $\lesssim 0.3^\circ$  at the lowest intensities and drops below $0.01^\circ$ for intensities $> 10^{14}$~W/cm$^2$. A detailed description of method and code, as well as numerical examples can be found in \cite{tao12,tsurf}.

The method used for the calculations labeled 'TDSE H 3' (I. Ivanov and A. Kheifets) has been monitored for convergence by changing the maximum angular momentum up to $L_{max}=80$, while the radial grid size was increased up to $r_{max}$=300 a.u. for calculations in both the length and the velocity gauge (with full agreement between the two).  The spectrum was obtained by projection onto the exact field-free continuum states of the H-atom after the end of the laser pulse. The step size of radial grid was $\delta r$=0.1 a.u. and the time-step was  $\delta t$=0.01 a.u.. Convergence was monitored by varying $\delta r$ down to 0.05 a.u. and the time-step $\delta t$ down to 0.005 a.u..

{The method used for two-electron systems has been described in detail in 
\cite{MCTDH} and is based on the Heidelberg MCTDH code, adapted to two electrons. It uses
time-dependent basis functions, variationally optimized to the electron dynamics, see \cite{MCTDH}. 
The electrons are treated in two dimensions each, with
basis functions set on the Cartesian grid with step-size  $\delta x$=0.2 a.u., covering $\pm 280$ a.u.
for each dimension. To achieve convergence 30 time-dependent basis functions per dimension are
used, leading to leading to $810,000$ total configurations propagated at each time step.}

\subsection{Acknowledgements}

We acknowledge stimulating discussions with U. Keller and A. Landsman.
J.K., O. S. and M. I. acknowledge  support of the EU Marie Curie ITN network CORINF. F. M. and O. S. acknowledge support of  the ERA-Chemistry grant, M. I. acknowledges support of the EPSRC Programme Grant, O.S.,L. T. and J.K. acknowledge support of the DFG grant SM 292/2-3. A.K. and I.I. acknowledge support of the Australian Research Council Grant DP120101805. A.Z. and A.S. acknowledge support by the DFG cluster of excellence ''Munich Center for Advanced Photonics (MAP)''. H. G. M. acknowledges the hospitality of the Max Born Institute.

$^{*}$ These authors have contributed equally.
 \subsection{Author Information}
 Correspondence should be addressed to O. S. (olga.smirnova@mbi-berlin.de)

%TC:endignore

\appendix

\section{The ARM theory applied to short circularly-polarized pulses}

\subsection{General expressions.}

The ARM approach has been described in detail in Refs.\cite{arm1,arm2,lisa,jivesh}, 
where it was originally developed for long laser pulses. Here we apply the same method for short pulses, 
taking into account effects of the pulse envelope. These effects are very important when 
considering the nearly-single cycle pulses required to perform atto-clock measurements.
%and provide a mapping between the ionization times and the angle-resolved photoelectron spectra

The ARM method yields the following expression for the photoelectron spectrum\cite{jivesh,lisa}:
\begin{equation}
	|a_{\mathbf{p}}(T)|^2 = |R_{\kappa l m}(\p)|^2 \ |e^{-i S(T,\p,t_s) }|^2
= |R_{\kappa l m}(\p)|^2 \ e^{2 \Imm S(T,\p,t_s) } ,
\label{spec}
\end{equation}
where $T \to \infty$.
The first term, $R_{\kappa l m}(\p)$, encodes the angular structure of the initial state. 
For the spherically symmetric ground state of hydrogen, which we focus on in 
this work, $R_{\kappa l m}(\p)$ does not impact the angle-resolved spectra. 
The action $S$ in the second term is comprised of three components,
\begin{equation}
	S(T,\p,t_s) = S_{V}(T,\p,t_s) + W_C(T,\p,t_s) - I_p t_s. \label{eq:S}
\end{equation}
The first of these is the so-called Volkov phase, the phase accumulated by the electron in the laser field only:
\bea
    	S_{V}(T,\p,t_s) = \frac{1}{2} \int_{t_s}^{T} dt \ [\p+\A(t)]^2 .  \label{Sv}
\eea
The second component is the phase accumulated due to the interaction of the departing electron with the core: 
\begin{equation}
	W_C(T,\p,t_s) =  \int_{t_s - i\kappa^{-2}}^{T} dt \ U(\r_s(\p,t,t_s)), \label{eq:WC}
\end{equation}
where $U(\r_s)$ is the potential of the atom or molecule evaluated along the electron's laser-driven quantum trajectory,  
%$\r_s(\p,t,t_s) =  \int_{t_s}^{t} dt' \ (\p + \A(t'))$.
\begin{equation}
	\r_s(\p,t,t_s) =  \int_{t_s}^{t} dt' \ (\p + \A(t')).
	\label{traj}
\end{equation}
%We stress that the integral Eq.\eqref{eq:WC} is convergent even for long-range potentials due to electron acceleration by the  strong laser field.
The third component comes from the evolution of the initial bound state, where  
$I_p$ is the ionization potential of this state.
Finally, $\kappa = \sqrt{2 I_p}$ in the lower limit of the integral in Eq.\eqref{eq:WC} comes from the 
matching of inner and outer region solutions. Each of the terms above is evaluated at 
the complex time $t_s = t_i + i \tau_T$, which is the solution to the saddle-point equation 
\begin{equation}
\left.\frac{\partial S(T,\p,t')}{\partial t'}\right|_{t_s} =
\left.\frac{\partial S_V(T,\p,t')}{\partial t'}\right|_{t_s} + \left.\frac{\partial W_C(T,\p,t')}{\partial t'}\right|_{t_s} -I_p
= 0
 \label{eq:saddle}
\end{equation}
The time $t_s$ defines the starting point of the electron trajectory, and the presence of the 
imaginary component in $t_s$ reflects the quantum nature of the electron's motion.

{The above expressions are obtained by solving the time-dependent Schroedinger equation
for the problem of strong-field ionization. The solution makes no
assumptions about the nature of the ionization process and is gauge-invariant \cite{arm1,arm2}.
The core part of the method relies on the rigorous R-matrix-type separation of 
coordinate space into inner and outer regions, with the wavefunction
transferred between the two using the Bloch operator -- a standard R-matrix technique. 
The approximations used by the analytical method are as follows:
(B1) The wavefunction in the inner region is approximated by the bound state from which ionization occurs. 
(B2) In the outer region, the method uses strong-field eikonal-Volkov 
states \cite{smirnovaCC} to describe the electron dynamics. The validity of 
the eikonal approximation sufficiently far from the core has been throroughly checked in \cite{smirnovaCC}. 
Its accuracy has been further verified for the case of delays in one-photon ionization measured 
by the attosecond streak-camera \cite{mivanov} -- the use of eikonal-Volkov states 
in the continuum yielded excellent agreement between analytical and ab-initio results.
(B3) The derivation also makes use of the saddle point 
method when evaluating the integrals that arise. Our ability to do so stems from the large action 
accumulated by the electron in the presence of a strong laser field.}

The key to our ability to reconstruct ionization times from photoelectron spectra comes from the fact 
that $t_i = \Ree[t_s]$, the real part of the saddlepoint solution above, is naturally interpreted as the time 
of ionization (see e.g.\cite{nirit} and references therein). Consequently, the analysis of ionization times in 
strong-field ionization is concerned with saddle-point times and the corrections to these times introduced by the core potential.
%These appear through $W_C$, as discussed below, especially its
%real part ${\rm Re} W_C$. 
%We note that the real-valued part of the  core contribution $W_C$
%Eq.(\ref{eq:WC}) to the action comes from the integration
%along the real time axis, $t>t_i={\rm Re}t_s$. If the electron
%coordinate $r_s({\bf p},t_i,t_s)$ exceeds the range of the core potential, then ${\rm Re} W_C=0$ (e.g. for a zero-range potential).

\subsection{Coulomb corrections to saddle-point times.}

It is instructive to start with the case where the electron-core interaction is negligible, which is appropriate e.g. for the short-range Yukawa potential.
If we neglect the term due to the Coulomb phase, Eq.\eqref{eq:saddle} reduces to 
\begin{equation}
\left.\frac{\partial S_V(T,\p,t')}{\partial t'}\right|_{t_s^0} - I_p = 0,
 \label{eq:saddle0}
\end{equation}
which can be easily solved with no approximations. 
For electron momentum ${\bf p}=\{p,\phi\}$,  we obtain
\begin{equation}
	\omega t_i^0({\bf p}) = \omega \Ree[t_s^0({\bf p})] 
	= \phi + \Delta \phi^\mathrm{env}({\bf p}),
	\label{eq:ti0}
\end{equation}
where $\phi$ is the angle at which the electron is detected 
and $\Delta \phi^\mathrm{env}$ is a small 
correction due to the shape of the pulse envelope. The sign of this correction depends on $\phi$: it is negative for $\phi>0$, 
positive for $\phi<0$, and zero for $\phi=0$. 
For sufficiently long pulses, $\Delta \phi^\mathrm{env}$ vanishes, and we are left with the simple 
mapping $t_i^0 = \phi/\omega$. That is, if we neglect electron-core interaction, we find that the 
angle of detection is orthogonal to the direction of the field at the moment of ionization.
This is fully consistent with the ab-initio numerical results obtained for the Yukawa potential, where
the majority of the electrons are detected orthogonal to the field direction at the peak
of the laser pulse.

In the absence of  electron-core interaction, the real part $t^0_i$ of the saddle point solution $t^0_s$ 
has a clear meaning. It corresponds to the so-called ionization time, since for all times  $t>t^0_i$ both the  
photoelectron spectrum  $|a^0_{\mathbf{p}}(t)|^2\propto e^{2 \Imm S_{SFA}(t,\p,t^0_s)}$ and the 
ionization probability $\int |a^0_{\mathbf{p}}(t)|^2 d\p$ remain constant: ionization is completed by  time $t^0_i$. 
Indeed, the imaginary component of the action $S_{SFA}(t,\p,t^0_s)$ is only accumulated while 
 integrating from $t^0_s$ to $t^0_i$ in Eq.(\ref{Sv}). This property of the integral in Eq.(\ref{Sv}) 
has prompted the perspective that tunnelling corresponds to  motion in imaginary time from 
$\Imm t^0_s$ to zero. $\Ree t^0_s=t^0_i$ then corresponds to the exit time: 
the time at which the electron leaves the tunnelling barrier.   Within this approach, 
in the absence of  electron-core interaction, tunnelling from the bound state starts and finishes at the
same real time $t^0_i$. 
% (Fig.1(d)) and ionization time directly maps
%into the electron detection angle $\phi$:
%\begin{equation}
%\w t^0_i=\phi+\phi^{env}.  \label{sfa_mapping}
%\end{equation}
%Here $\phi^{env}$ is a small correction to the standard 
%mapping $\w t^0_i=\phi$ due to the rapidly changing envelope of the single cycle pulse;
%Eq. (\ref{sfa_mapping}) is the real part of the exact solution of Eq. (\ref{saddle0}).
%Thus, the electron detection angle measures the time
%when  ionization was completed.
This result is fully consistent with the ab-initio calculations for the Yukawa 
potential presented in the main body of the paper. We note that the potential 
used in the numerical calculations has a single bound s-state, which ensures 
that scattering phases in all ionization channels other than the s-channel are equal to zero.

With this result in mind, we can express the full solution to Eq.\eqref{eq:saddle} as
\begin{equation}
	t_s =t_s^{0}+ \Delta t_s^C,
\end{equation}
where $t^0_s$ is the solution of the Coulomb-free saddlepoint equation (Eq.\eqref{eq:saddle0}) and
 $\Delta t_s^C$ is the correction due to the electron-core interaction. 
Expanding Eq.\eqref{eq:saddle} in a Taylor series around $t^0_s$ and keeping all terms up to first 
order in the electron-core interaction $W_C$, we obtain
\begin{equation}
\Delta t_s^C=-{\frac{\partial W_C(T,\p,t^0_s)}{\partial t^0_s}}\left({\frac{ \partial^2 S_{V}(T,\p,t_s^0)} {(\partial t_s^0)^2}}\right)^{-1}.
 \label{eq:DtsC01}
\end{equation}
Next, we note that Eq.\eqref{eq:saddle0} establishes a functional dependence $t_s^0=t^0_s(I_p)$. 
With this in mind, differentiating both sides of Eq.\eqref{eq:saddle0} with respect to $I_p$, we have
\begin{equation}
	\left({\frac{ \partial^2 S_{V}(T,\p,t_s^0)}
	{(\partial t_s^0)^2}}\right)^{-1}
	=\frac{d t^0_s}{d I_p}. \label{eq:Ipts}
\end{equation}
Combining Eq.\eqref{eq:DtsC01} and \eqref{eq:Ipts}, we obtain
\begin{equation}
	\Delta t_s^C=-\frac{\partial W_C(T,\p,t^0_s)}{\partial t^0_s}\frac{ d t^0_s} {d I_p}.
	\label{eq:DtsC}
\end{equation}

To derive a practical way of calculating $ \Delta t_s^C$, we 
recall that $W_C(T,\p,t^0_s)$ (Eq.\eqref{eq:WC}) depends on $I_p$ only 
via $t_s^0=t_s^0(I_p)$ and $\kappa=\sqrt{2 I_p}$ in the lower limit of the integral. 
The full derivative of $W_C(T,\p,t^0_s)$ with respect to $I_p$ can therefore be expressed as
\begin{equation}
        \frac{d W_C(T,\p,t^0_s)}{d I_p}=\frac{ \partial W_C(T,\p,t^0_s)}
        {\partial t^0_s}\frac{ d t^0_s} {d I_p}+\frac{ \partial W_C(T,\p,t^0_s)}
        {\partial \kappa}\frac{d \kappa} {d I_p}
        . \label{eq:ful_d}
\end{equation}

Thus, we can evaluate $\Delta t_s^C$ by differentiating $W_C(T,\p,t_s^0)$ with 
respect to $I_p$, while keeping $\kappa$ in the lower limit of the integral constant: 
\begin{equation}
	\Delta t_s^C = -\frac{d W_C(T,\p,t_s^0)|_{\kappa=const}}{d I_p}.
%	%\ = \ -\frac{d}{d I_p} W_C(T,\p,t_s^0)|_{\kappa=const}.
	\label{eq:DtsC2}
\end{equation}
%For the real part, this simplifies to
%\begin{equation}
%	\Delta t_i^C = \Ree \Delta t_s^C = -\frac{d \Ree [W_C(T,\p,t_s^0)]}{d I_p}. 
%	\label{eq:DtiC}
%\end{equation}
%The above quantity, $\Delta t_i^C = \Ree \Delta t_s^C$ and represents a Wigner-Smith-like ionization delay for strong-field ionization. Essentially, it is the time by which an electron moving in a long range potential is delayed compared to an electron which does not interact with the core after ionization.
%
%We note that, in the one-photon case, the ionization  delay is given by the derivative of the phase of the photoionization matrix element with respect to the photoelectron energy. In contrast to scattering problems, this phase is finite and converges even for the long range potential due to the finite extent of the bound state. Thus, photoionization delay is finite for both one-photon and strong field cases (see discussion above after Eq.\eqref{eq:WC}.)

Recalling that the real part of the saddlepoint solution, $t_i=\Ree[t_s]$, represents the time of ionization, 
our analysis has established a mapping 
between the angle and momentum at which the electron is detected and its ionization time $t_i=\Ree[t_s]$:
\begin{equation}
	t_i(\phi,p) = \frac{\phi}{\omega} + \Delta t_i^\mathrm{env}(\phi,p)
+ \Delta t_i^C(\phi,p),
\label{eq:ti}
\end{equation}
where $\Delta t_i^\mathrm{env}$ is obtained by solving Eq.\eqref{eq:saddle0}, and $\Delta t_i^C= \Ree \Delta t_s^C$ 
is calculated using Eq.\eqref{eq:DtsC2} above.

%We note that if the electron
%coordinate $r_s({\bf p},t_i,t_s)$ at the moment
%$t_i={\rm Re}t_s$
%exceeds the range of the core potential, then ${\rm Re} W_C=0$, leading to no correction of the real part of the saddle
%point. As the range of the potential is gradually increased,
%the phase 
%${\rm Re} W_C=0$ will start to deviate from zero, leading
%to non-zero $\Delta t_i^C$.

Finally, we discuss the dependence of $W_C$ (SM, Eq.(\ref{eq:WC}))
on the range $a$ of the short-range Yukawa potential. The integral 
Eq.(\ref{eq:WC}) is calculated from $t_s$ down to the real time axis, $t_i=\Ree(t_s)$, 
and then along the real time axis. In the classically forbidden region, 
between $t_s$ and $t_i$, the differential $dt$ is purely imaginary and thus, 
for real-valued $U(r)$, no real-valued contribution to $W_C$ is accumulated.
(The potential $U$ evaluated along the complex trajectory 
remains real-valued in the classically 
forbidden region between $t_s$ and $t_i$ as long as
$\Ree r_s > \Imm r_s$. This condition is met for the peak of
the electron distribution.) The real-valued contribution to $W_C$ accumulates at $t>t_i$.  If $r_s(p,t_i,t_s)$ exceeds the 
range of the potential, then $U(r_s(p,t\geq t_i,t_s)=0$, and
$Re W_C=0$, leading to no correction to the real part of the saddle point. 
Obviously, the same applies to the zero-range potential. As the range $a$ of the potential becomes comparable
to $r_s(p,t_i,t_s)$, $\Ree W_C$ will start to deviate from
zero, leading to non-zero $\Delta t_i$.

\section{Origins of the negative ionization times.}\label{sect:NegativeTimes}

Fig. 5 in the main body of the paper shows that for the parameters of the numerical experiments, 
ionization times become negative at intensities $I>1.5 \times 10^{14}$W/cm$^2$, 
i.e. for electric field strengths $F_L>0.0465$ a.u.
One of the possible explanations for this trend could be the depletion of 
the ground state. Indeed, if the ground state population is 
significantly reduced before the peak of the laser field, the maximum ionization 
signal may precede the field maximum.
Thus, we first check the role of the depletion of the ground state  in this trend.

\subsection{Effect of ground state depletion on ionization times.}\label{sect:NegativeTimes1}

Fig. 5 (b) (see main text)  shows the population of the ground state after
the end of the pulse, calculated using the `TDSE H1' approach (F. Morales and H. Muller).
The population is obtained by projecting the wavefunction of the system on the ground state after
the end of the pulse.
The depletion of the ground state remains very small until $I=2 \times 10^{14}$W/cm$^2$, and is below $10^{-3}$ when
negative ionization times become apparent. However, 
these times (Fig. 5 (a,c)) are also small. 
At higher intensities, ionization reaches a few percent and depletion of the ground state population must contribute to the
increasing values of negative ionization times. 
To quantify this effect,  we take advantage of the attoclock mapping
between the ionization times $t_i$ and the electron deflection angle $\theta$.
We also take into account that both the depletion of the ground state and the negative ionization times are very small.
  
Consider the photoelectron distribution integrated over energy, $P(\phi)$, and
let us denote by $P_0(\phi)$  the photoelectron distribution in 
the absence of depletion. According to the ARM theory, their
relationship is $P(\phi)=P_0(\phi)W_g(t_i(\phi))$, where $W_g(t_i(\phi))$
is the ground state population at the moment of ionization
$t_i(\phi)$, which is mapped to the detection angle $\phi$. 
If ground state depletion is the sole origin of the negative ionization times, 
then $P_0(\phi)$ should have its maximum at the angle which corresponds to 
an ionization time of zero: that is, at the angle $\phi=\theta_0$ which satisfies $t_i(\theta_0)=0$. 
The decreasing function $W_g(t_i)$ will then skew this distribution towards $t_i<0$.
Note that, according to the attoclock principle, 
$\phi-\theta_0=\omega t_i$, where $\omega$ is the laser frequency.

For small ionization times near the peak of 
the laser field $t=0$,  we have $W_g(t_i)=W_g(0)[1-\Gamma_0 t_i]$, 
where $\Gamma_0=\Gamma(t=0)$ is ionization rate at the peak of the field, and $W_g(0)$ is the ground state population at the peak of the field.
To determine how the decreasing value of $W_g(t_i)$ shifts the maximum
of $P(\phi)$,  we  expand $P(\phi)$ in a Taylor series around $\theta_0$, the angle at which $P_0(\phi)$ is maximized:\begin{align}
	P(\phi) &=P_0(\phi)W_g(0) [1- \Gamma_0 t_i]
      \label{Taylor} \\
	&=\left[P_0(\theta_0)-\frac{1}{2}P_0''(\phi-\theta_0)^2\right]W_g(0)\left[ 1-\Gamma_0 t_i\right].
 \label{Taylor1}
\end{align}
Here we have used the fact that $P_0(\phi)$ has its maximum at $\phi=\theta_0$ and hence $P_0'(\theta_0)=0$. 
Using $\phi-\theta_0=\omega t_i$ we obtain:
\begin{align}
	P(\phi) &=P_0(\phi)W_g(0) [1- \Gamma_0 t_i]
      \label{Taylor2} \\
	&=\left[P_0(\theta_0)-\frac{1}{2}P_0''\omega^2 t_i^2\right]W_g(0)\left[ 1-\Gamma_0 t_i\right].
 \label{Taylor3}
\end{align}
We can now find the maximum of the function $P(\phi)$ by solving the equation $dP(\theta)/d\phi=0$, or equivalently $dP/dt_i=0$, given the linear dependence 
$\phi-\theta_0=\omega t_i$. 
%
%Evaluating $\frac{dP(\theta(t_i))}{d\theta(t_i)}=0$ explicitly we obtain the following
%equation for $t_i$:
%\begin{eqnarray}
%     \frac{3}{2}\Gamma_0 P_0^{''}\omega^2t_i^2-P_{\theta,\theta}''\omega^2 t_i=\Gamma_0 P_0(\theta_0)
% \label{quadraticeq}
%\end{eqnarray}
For small time shifts $t_i$, where $\Gamma_0 t_i\ll 1$, this yields 
\begin{eqnarray}
    \omega t_i=-\frac{P_0(\theta_0)}{P_{0}^{''}(\theta_0)}W_g(0)
	\Gamma_0 \frac{1}{\omega}
%	=
%	-\frac{P_0(\theta_0)}{P_{0}^{''}}
%	\Gamma_0 \frac{1}{\omega},
 \label{solution}
\end{eqnarray}
In our numerical simulations, $\Gamma_0 t_i\leq 4\times 10^{-3}$ for all intensities
shown in Fig.5. We can now use our ab-initio photoelectron distributions to extract 
the required quantities, the ratio $P_0/P_0''$ 
and the ionization rate $\Gamma_0$. 

\begin{figure}
\includegraphics[width=5in,angle=0]{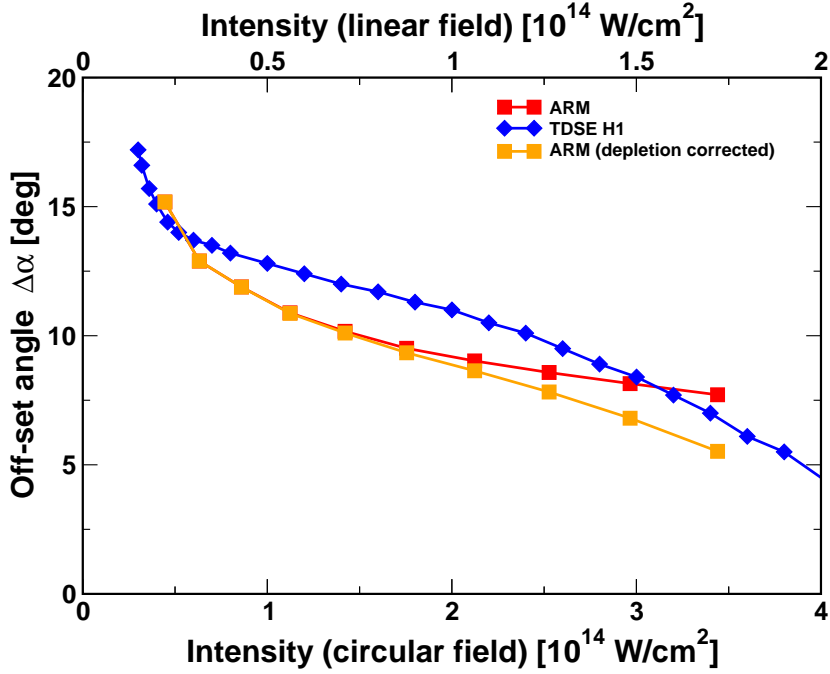}
\caption{ Offset angles shown as a function of intensity. Red squares represent the angles 
obtained using the ARM method with depletion neglected. Orange squares show the same ARM results, corrected for depletion according to Eq.\eqref{final_omegati} using 
numerically derived values for $W_{\rm Loss}$ (the depletion of the ground state) and  $\Delta$ (the width of the photoelectron distribution). 
Blue diamonds show the offset angles extracted from ab-initio photoelectron spectra calculated using method TDSE H2. 
At higher intensities, the depletion-corrected ARM angles are better able to reproduce the trend we see in the numerical results.
}  \label{FigSM1}
\end{figure}

First, we integrate each ab-initio energy- and angle- resolved photoelectron 
distribution over energy to obtain an angle-resolved electron yield, to which we fit a Gaussian: 
 \begin{eqnarray}
   P(\phi)=C\exp\left[-\frac{(\phi-\theta_{\rm max})^2}{2\Delta^2}\right],
 \label{solution1}
 \end{eqnarray}
where $\theta_{\rm max}$ corresponds to the maximal photoelectron signal. 
Again, making use of $\Gamma_0 t_i\ll 1$,  and using Eq.(\ref{Taylor}), we have
 \begin{eqnarray} 
\frac{P_0(\theta_0)}{P_{0}^{''}(\theta_0)}= \frac{P(\theta_{\rm max})}{P^{''}(\theta_{\rm max})}=\Delta^2,
	\label{ratio}
 \end{eqnarray}
where the relative error in the first equality 
in the above equation 
 is $\Gamma_0 t_i/2\leq 2 \times 10^{-3}$. Eq.(\ref{ratio}) 
yields the following expression for $\omega t_i$:
 \begin{eqnarray}
 \omega t_i=-\frac{\Delta^2}{\omega}\Gamma_0 W_g(0).
 \end{eqnarray}
Finally, we note that the attoclock mapping $\theta(t_i)$ reflects the sub-cycle dependence of the ionization rate,
which follows the same Gaussian-shaped distribution as the photoelectron spectrum, 
 \begin{eqnarray}
  \Gamma(\theta)=\Gamma_0 \exp\left[-\frac{(\phi-\theta_{0})^2}{2\Delta ^2}\right]
=\Gamma_0 \exp\left[-\frac{(\omega t_i)^2}{2\Delta^2}\right]\ \ \ .
  \label{gamma_subcycle}
  \end{eqnarray}
Here, we explicitly use the assumption that $\Gamma$ is maximized at the peak of the field, i.e. depletion is the only reason for negative ionization times ($\theta_0$ corresponds to the peak of the field $t=0$). 
For a circularly polarized field, we can obtain the total ionization yield $W_i$  
by integrating $\Gamma(\phi)$ over all angles $\phi$ or, equivalently, over all $t_i$. This yields
 \begin{eqnarray}
 W_i=\Gamma_0 \sqrt{2\pi}\frac{\Delta}{\omega}.
\label{population_loss}
  \end{eqnarray}
This equation can be used to obtain the ionization rate $\Gamma_0$ from the total ionization signal or from the depletion of the 
ground state $W_{\rm Loss}\simeq W_i$ (in all our numerical simulations these quantities differ by less than 1\%):
\begin{eqnarray}
\Gamma_0= \frac{W_{\rm Loss}}{\sqrt{2\pi}} \frac{\omega}{\Delta } ,
\end{eqnarray}
Thus, numerically finding the depletion of the ground state and the angular width of the 
photoelectron distribution, we have all the quantities necessary to obtain the 
negative ionization times due to ground state depletion: 
\begin{eqnarray}
  \omega t_i=- W_g(0) W_{\rm Loss}\frac{\Delta }{\sqrt{2\pi}}.
  \label{final_omegati}
  \end{eqnarray}
%  \begin{figure}
%%\includegraphics[width=5in,angle=0]{figure-1.pdf}
%\caption{The importance of ground state depletion. (a) Population of the ground state of the H-atom after the end of the laser pulse, as a function of
%the circular pulse peak intensity, for $\phi_{\rm CEP}=0$. (b) Ionization times (green triangles) 
%obtained from the photo-electron spectra, as a function of the circular pulse peak intensity. 
%The peak of the field corresponds to
%$t=0$, thus zero ionization time corresponds to ionization at the peak of the laser field. Red circles 
%show the effect of the ground state depletion on ionization times, calculated using Eq.(\ref{final_omegati})
%with $\Delta$ and $W_i$ obtained from ab-initio simulations. 
%}  \label{FigSM1}
%\end{figure}
The equation above was used to obtain the depletion-corrected ionization times in Fig. 5 (c) 
of the main text (red circles). We have used $W_g(0)=1$, which provides an upper 
bound for the effect: at the highest intensity shown, substituting $W_g(0)=1$ leads to 
an error below 0.25$^\circ$ (at lower intensities, the error is considerably less). 
The same equation can also be used to correct the ARM offset angles as shown in Fig. 3 (orange squares) of the main text. The results of doing so are also shown in  Fig.\ref{FigSM1}. 
Figure 5 (c) of the main text indicates that although depletion is indeed partially responsible for 
the appearance of  negative times, it cannot explain them fully. This suggests that there could be additional effects at play.

%\begin{eqnarray}
%      \mathbf{A}_L(t) = -\frac{A_0}{\sqrt{1+\epsilon^2}} \ \cos^4(\w t/4) \ (\epsilon 
%	\cos(\w t+\phi_{\rm CEP}) \ \mathbf{\hat{x}} +  
%	\sin(\w t+\phi_{\rm CEP}) \ \mathbf{\hat{y}}).
% \label{pulse}
%\end{eqnarray}
 
%However, Fig.SM2 shows that depletion of the ground state population is not the only contributing factor.
\subsection{Additional effects contributing to negative ionization times.}
\label{sect:NegativeTimesP}

One possible additional effect is suggested by the ARM theory. Within the ARM picture, we find that, 
for a long-range potential, 
the total ionization yield continues to change even after the electron has emerged from the barrier. 
That is, ionization is not fully completed during the tunnelling step. This signifies the post-tunnelling step in ionization.  
Such a situation, in fact, is familiar in the context of ionization in linearly polarized fields: 
in `frustrated tunnelling' the liberated electron returns to its parent ion and can become trapped in excited states. 
As discussed below,our numerical tests suggest that the negative ionization times we reconstruct could 
point to a similar effect for circularly polarized fields, where such an effect becomes possible for nearly single cycle pulses.

If ionization yield depended on the tunnelling step only, we would expect the 
highest yield to be associated with the peak of the field, since this corresponds to the thinnest tunnelling barrier. 
That is, once effects of depletion have been taken into account and in the absence of any 
tunnelling delays, we would expect an optimal ionization time $t_i = 0$. However, since the `post-tunnelling' step 
also contributes to the yield, \emph{both} steps must be optimized at the maximum of the photoelectron distribution. 
ARM shows that the post-tunnelling step favours ionization at earlier times $t_i < 0$. 
This could explain the negative ionization times we reconstruct in Figure 5 of the main text. 
We would expect that features associated with the post-tunnelling step would become more pronounced at higher 
intensities, where deviation from the maximum of the field has a relatively smaller impact on the efficiency of 
the tunnelling process. Indeed, this is where we observe the largest negative ionization times. It is 
possible to test this idea further by investigating the CEP dependence of ground state depletion and
 total ionization yield in elliptical fields. Such tests are discussed below.

\subsection{Numerical tests of additional effects:  CEP control in elliptical fields.}
\label{sect:NegativeTimes2}

Changing the ellipticity of the laser pulse from perfectly circular
to nearly circular adds an important control parameter: the carrier-envelope phase (CEP) of the 
ultrashort pulse, $\phi_{\rm CEP}$. 
Changing $\phi_{\rm CEP}$ varies the direction of the electric field vector at the peak of
the intensity envelope. 
For perfectly circular polarization, CEP does not affect the peak value of the electric
field nor the shape the photoelectron spectra -- it merely rotates the spectra by an angle equal to $\phi_{\rm CEP}$. 
For $\epsilon<1$ on the other hand, assuming the orientation of the polarization ellipse is fixed, $\phi_{\rm CEP}$ controls the peak value of the
electric field strength.

\begin{figure}
\includegraphics[width=5in,angle=0]{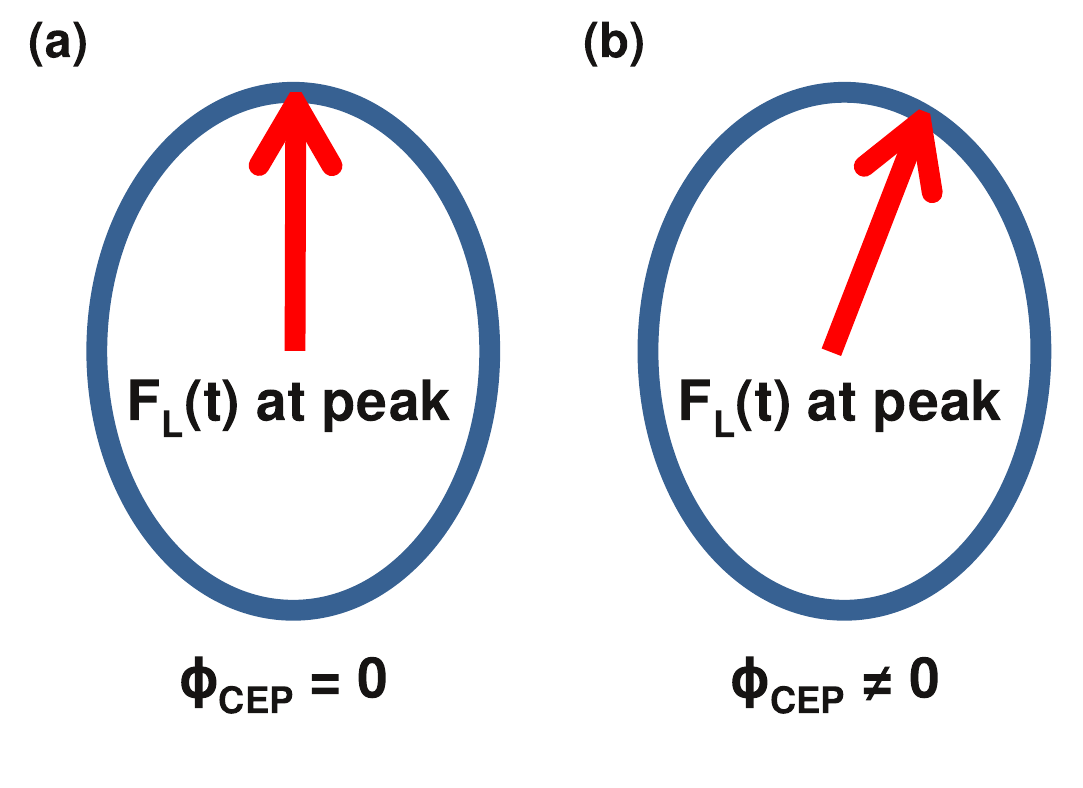}
\caption{ A schematic illustration of the way in 
which carrier-envelope phase (CEP) controls the peak field strength for elliptically polarized pulses. 
(a) $\phi_{\rm CEP}=0$ maximizes the peak field strength, (b) For $\phi_{\rm CEP}\neq 0$, the peak field strength is reduced.
}  \label{FigSM2}
\end{figure}

In the simulations carried out, the laser field was defined by $\F_L(t)=-{\partial\mathbf{A}_L(t)}/{\partial t}$, where
\begin{eqnarray}
      \mathbf{A}_L(t) = -\frac{A_0}{\sqrt{1+\epsilon^2}} \ \cos^4(\w t/4) \ (\epsilon 
	\cos(\w t+\phi_{\rm CEP}) \ \mathbf{\hat{x}} +  
	\sin(\w t+\phi_{\rm CEP}) \ \mathbf{\hat{y}})  \ \ \ ,
 \label{pulse}
\end{eqnarray}
and $\epsilon=0.85$ . 
The electric field polarization ellipse has its major axis along the $y-$direction in this case.
For $\phi_{\rm CEP}=0$, the peak of the electric field envelope coincides with the orientation of the 
major axis of the polarization ellipse, ensuring the maximum value of the electric field, see Figure \ref{FigSM2} (a).
Changing $\phi_{\rm CEP}$ changes the direction of the electric field vector at the peak of
the intensity envelope and reduces the peak field strength, see  Figure \ref{FigSM2} (b).

The standard tunnelling perspective, expressed in the attoclock postulates A1-A3 (see main text), states that tunnelling
is optimized at  the maximum of the electric field, when the tunnelling barrier is at its thinnest. 
Based on this perspective, we would expect
$\phi_{\rm CEP}=0$ to lead to the maximal total ionization yield, which is the commonly made assumption 
when interpreting attoclock experiments. If, however, negative ionization times reflect the 
fact that ionization is not completed during the tunnelling step and the electron's motion must be 
optimized throughout both the classically forbidden and classically allowed regions, then $\phi_{\rm CEP}=0$ may 
not be optimal, depending on the relative contribution from the classically allowed region. Indeed, $\phi_{\rm CEP}=0$ will only optimize 
the tunnelling step, since it maximizes the instantaneous value of the electric field. 

In light of this, we have performed a CEP-scan of angle- and energy-photoelectron spectra across a range of laser intensities from $I=1\times 10^{14}$W/cm$^2$ to
$I=3 \times 10^{14}$W/cm$^2$ for a fixed carrier wavelength of $\lambda$=800 nm,
with $\epsilon=0.85$. We find that at lower intensities, $\phi_{\rm CEP}=0$ is indeed optimal. It yields the maximum 
total ionization signal and the highest peak in the angle-and-energy resolved photoelectron distribution.
%The situation changes exactly at the same intensity where the ionization times become negative: the optimal  $\phi_{\rm CEP}$ starts to deviate from zero.
However, at higher intensities we find that the optimal CEP starts to deviate from zero. 
This change occurs at the same intensity at which ionization times became negative.

\begin{figure}
\includegraphics[width=5in,angle=0]{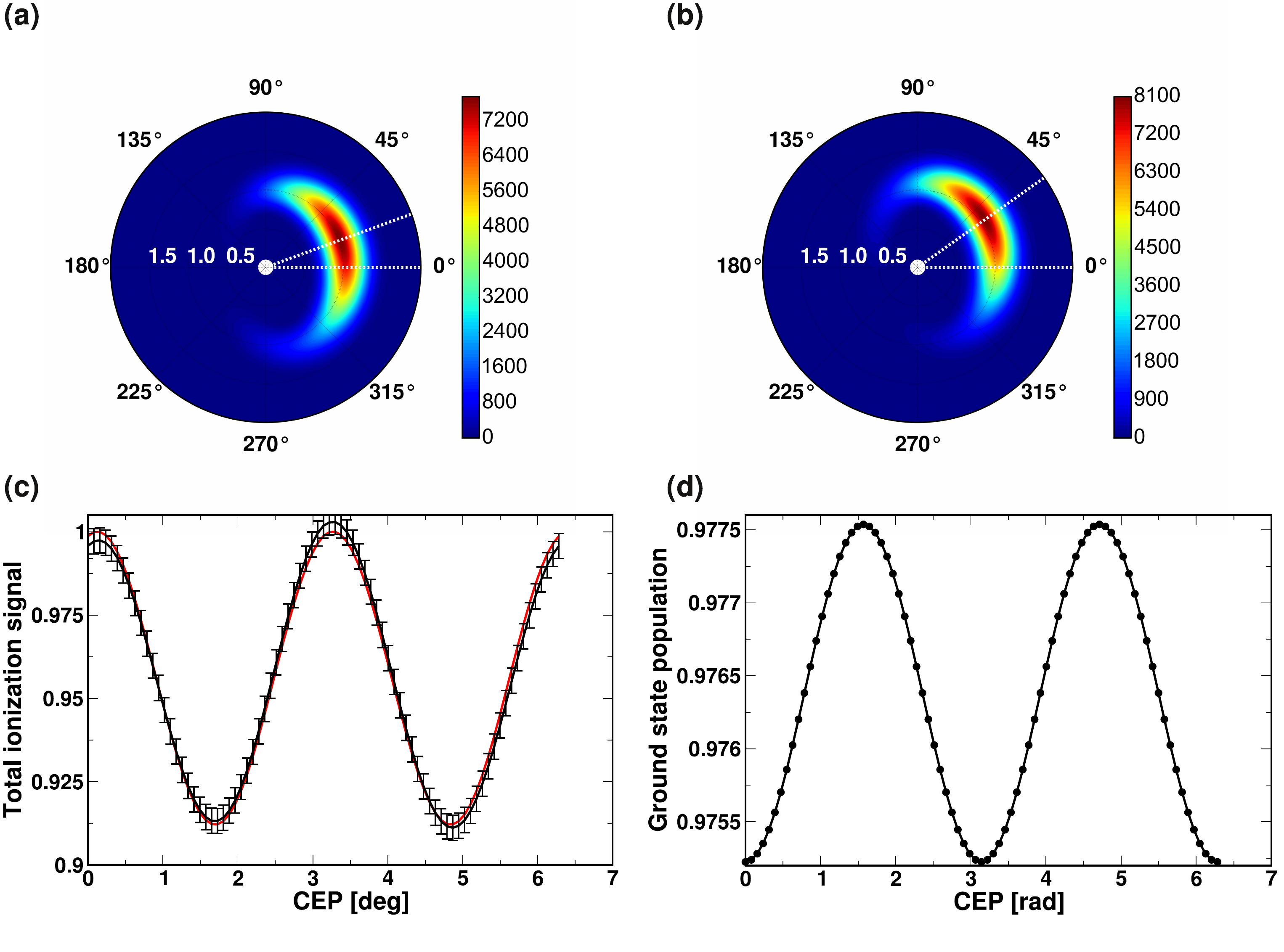}
\caption{  Sensitivity of strong-field ionization to the carrier-envelope phase $\phi_{\rm CEP}$ for elliptically polarized fields. 
The calculations were done for $\lambda=800$nm, $F_L=0.06$ a.u and $\epsilon=0.85$ using method TDSE H1.
Panels (a,b) show angle- and energy- resolved spectra for
$\phi_{\rm CEP}=0^\circ$ and $\phi_{\rm CEP}=20^{\circ}$ respectively. Although the maximum value of the instantaneous electric field is lower in the second case, the overall strength of the photoelectron signal is higher. Panel (c) shows the total ionization signal as a function of $\phi_\mathrm{CEP}$, integrated over all electron energies and angles, and normalized to 1 (black curve). The error bars reflect the accuracy of the numerical calculations. The red curve shows the optimal sinusoidal fit to the numerical data. This fit shows that the optimal $\phi_\mathrm{CEP}$, which maximizes the total ionization signal, is $0.25 \pm 0.15$ rad. Panel (d) shows the depletion of the ground state as a function of $\phi_{\rm CEP}$. 
}  
\label{FigSM3}
\end{figure}

The results are presented in the  Figure \ref{FigSM3}. Panels (a,b) show the 
angle- and energy- resolved spectra for a laser pulse with $F_L=0.06$ a.u., 
(a) $\phi_{\rm CEP}\simeq 0^{\circ}$ and (b) $\phi_{\rm CEP}\simeq 20^{\circ}$.  We find that the 
peak in the photoelectron spectrum is slightly higher for $\phi_{\rm CEP}\simeq 20^{\circ}$.
Panel (c) shows the total ionization yield as a function of $\phi_{\rm CEP}$. 
This curve reaches its maximum near $\phi_{\rm CEP}\simeq 20^{\circ}$, 
when the maximum electric field is lower than for $\phi_{\rm CEP}=0^\circ$. 
Panel (d) shows the depletion of the ground state vs $\phi_{\rm CEP}$, which is indeed maximized at $\phi_{\rm CEP}=0^\circ$.

Given the high nonlinearity of the tunnelling process, if the tunnelling
step was the only deciding factor in determining the ionization probability, the total ionization yield would have been optimized for $\phi_{\rm CEP}=0$. The deviation of the optimal $\phi_{\rm CEP}$ from zero in panel (c), while depletion is indeed maximized at $\phi_{\rm CEP}=0$ (panel (d)), is consistent with the idea that the ionization process is not completed by the end of the tunnelling step. 
After the tunnelling step is completed, the electron may again be trapped by the ion, permanently or transiently (in which case it would again be re-released by the atom). In both cases, the shape of the photoelectron distribution at the detector 
(the attoclock observable) is affected since a fraction of photoelectrons either never reach the detector or reach it with different momenta due to transient trapping. We find that, for the parameters of these numerical experiments, about 
0.01\% of the electrons are trapped in Rydberg states after the end of the laser pulse, corresponding to about 1\% of the number of 
the electrons that have left the ground state. The fact that this effect is nevertheless visible in the reconstructed ionization times shows that these times can serve as a highly sensitive probe of ionization dynamics.

% Thus, the Wigner-Smith-like time-delay $\Delta t_C$, which we have derived here for the strong-field regime, represents not just the effective deflection of electron trajectories compared to the zero-range potential, but also the ongoing nature of the ionization process. In Ref.\cite{lisa} we have shown
% that after exiting the tunnelling barrier the population of quantum trajectories indeed 
% continues to change, so that the total continuum population can experience a transient minimum after the tunnelling step is completed.

%Our results demonstrate that, for nearly single-cycle pulses, the common assumption in identifying the origin of the deflection angle in the CEP-averaged photoelectron spectra must be carefully tested and benchmarked for the specific parameters of the experiment if one aims to achieve 10-asec accuracy. This must include the characterization of ellipticity $\epsilon$ for all spectral components of the broad-band pulse, the characterization of the pulse envelope (preferably at the sub-cycle level),  and intensity calibration. Such benchmarking and calibration will also necessarily require ab-initio simulations. In contrast, the ab-initio simulations for the hydrogen atom we have presented here are free from all of the above uncertainties, since all the field parameters are under our complete control and are precisely known.

\section{Multielectron effects}

\begin{figure}
\includegraphics[width=5in,angle=0]{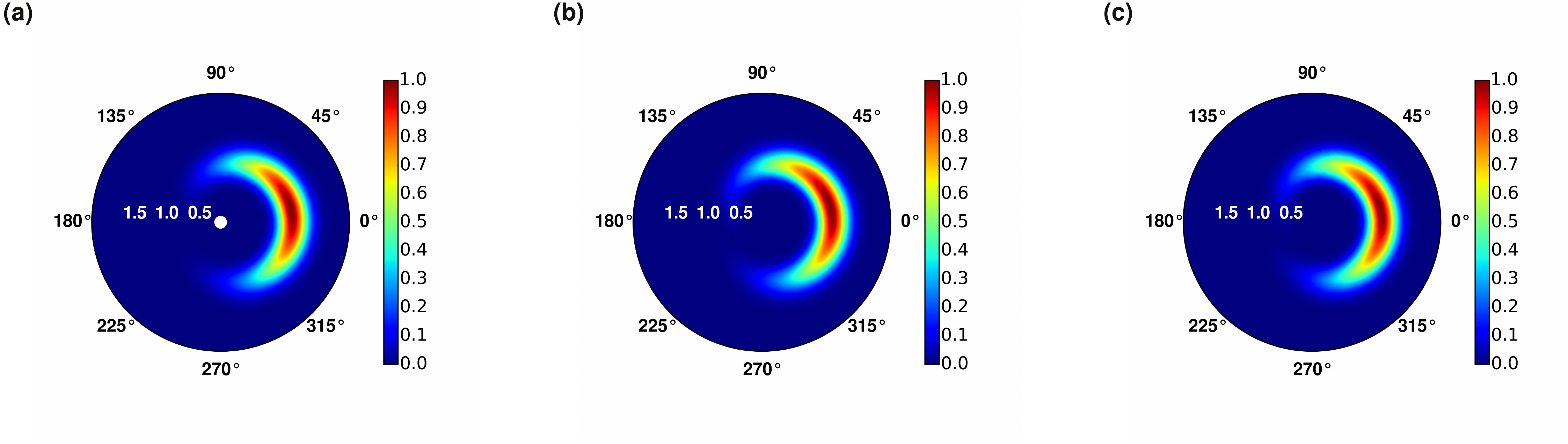}
\caption{ Comparison of attoclock 
observables for one-electron and two-electron systems.  
Panels (a,b,c) show angle- and energy- resolved spectra for
the hydrogen atom (a), a model 2D hydrogen atom with 
the same ionization potential as hydrogen (b),
and a model 4D two-electron system with the same ionization potential as
hydrogen (c). The calculations were done for $\lambda=800$nm, $F_L=0.055$ a.u,  $\epsilon=1$,
 and the same pulse shape as in the main text. 
}  \label{FigSM4}
\end{figure}

{To demonstrate how the calibration of the attoclock measurements for the one-electron system can be
used to identify multi-electron contributions, we have performed ab-initio calculations for several
model two-electron systems, where the system parameters were adjusted to change the role
of two-electron processes. The motion of both electrons was restricted to 2D each, making it
a 4D system. We used the soft-core
Coulomb potential for the electron-electron and electron-core interactions, }
\begin{eqnarray}
   V(Z,\r_1,\r_2)= -\frac{Z}{\sqrt{\r_1^2+a^2}} -\frac{Z}{\sqrt{\r_2^2+a^2}}
	+\frac{1}{\sqrt{|\r_1-\r_2|^2+a^2}}
\ \ \ ,
 \label{SoftCore}
\end{eqnarray}
{First, we consider a two-electron atom with the doubly charged core, Z=+2, with $a=1.12$ a.u. to yield
the same ionization potential as the hydrogen atom, $I_p=0.5$ a.u. The calculations are done using the
method described  in 
\cite{MCTDH}. It  is based on the Heidelberg MCTDH code, adapted to two electrons. It uses
time-dependent basis functions, variationally optimized 'on-the-fly' to the electron dynamics, see Ref.\cite{MCTDH}. 
The electrons are treated in two dimensions each, with
basis functions set on the Cartesian grid with step-size  $\delta x$=0.2 a.u., covering $\pm 280$ a.u.
for each dimension. To achieve convergence, 30 time-dependent basis functions per dimension are
used, leading to $810,000$ total configurations propagated at each time step.
We have also performed calculations
for the 2D H-atom, setting Z=1 and a=0.8 a.u., again to yield
the same ionization potential as hydrogen atom, $I_p=0.5$ a.u.
%±280 a.u., and 25-30  basis functions per degree of freedom 
%leads to 130 − 230 hours of computational time per one intensity point.
The sample photo-electron spectra for the three systems are shown in the  
Fig. \ref{FigSM4}, for $I=2.1 \times 10^{14}$W/cm$^2$. 
Comparing the spectra, which are extremely similar, demonstrates that single-electron effects are
indeed dominant and that the ionization potential and the long-range interaction with
the core determine the attoclock offset angle as long as the ionization yields are close.}

%For low ionization probabilities, the 2D and real 3D
%H-atoms yield similar offset angles $\theta$, within the numerical error bars. 
{Both 2D and 4D 
soft-core systems ionize more easily than real 3D hydrogen, with ionization 
reaching 10\% at $I=2.3 \times 10^{14}$W/cm$^2$. However, the ionization yield is virtually identical in both 
the one-electron and the two-electron soft-core systems, making their comparison 
straightforward and unambigous. The difference in the deflection angle
between the 2D and 4D system is shown in Fig.5(d) of the main text. It 
deviates slightly from zero, with
a small {\it negative} $\delta\theta$ for the two-electron system relative to the one-electron case. }

%Population left in the 
%excited states after the end of the laser pulse is also very small, well below 1\%, 
%consistent with adiabatic response of the core.
{The fact that we do not see any substantial positive delays due to two-electron effects in this case is not surprising. Indeed, for this model two-electron system, the first excited state of 
the ion is 0.47 a.u. above the ground ionic state, i.e. at almost twice the first ionization potential. 
Therefore, two-electron excitations in the IR laser field are very unlikely.
The small reduction of the deflection angle is fully consisent with the adiabatic polarization
of the ionic core by the laser field, which reduces the attraction 
of the outgoing electron to the ion. }

{Next, we have explored the conditions under which two-electron excitations in the IR field may play
a bigger role. To this end, we have considered a substantially less 'rigid' 
model 4D two-electron system with $a=2.925$ a.u.  and $Z=+3$. 
The substantially softer interaction reduces
excitation energies in the system, while $Z=+3$ maintains the same first
ionization potential $I_p=0.5$ a.u. as before. Now the first excited
ionic state is only 0.213 a.u. above the ionization threshold. Once again, we have paired
this model system with a one-electron 2D system that has the same long-range
core potential (Z=+2), the same $I_p=0.5$ a.u. ($a=2.7$ a.u.), and a very similar ionization yield vs laser intensity.
When single ionization approaches 50\%, electron-electron correlation leads
to a substantial positive $\delta \theta$ for the two-electron system compared to the one-electron
system, as shown in the Fig.5(d) of the main text.}

\begin{figure}
\includegraphics[width=5in,angle=0]{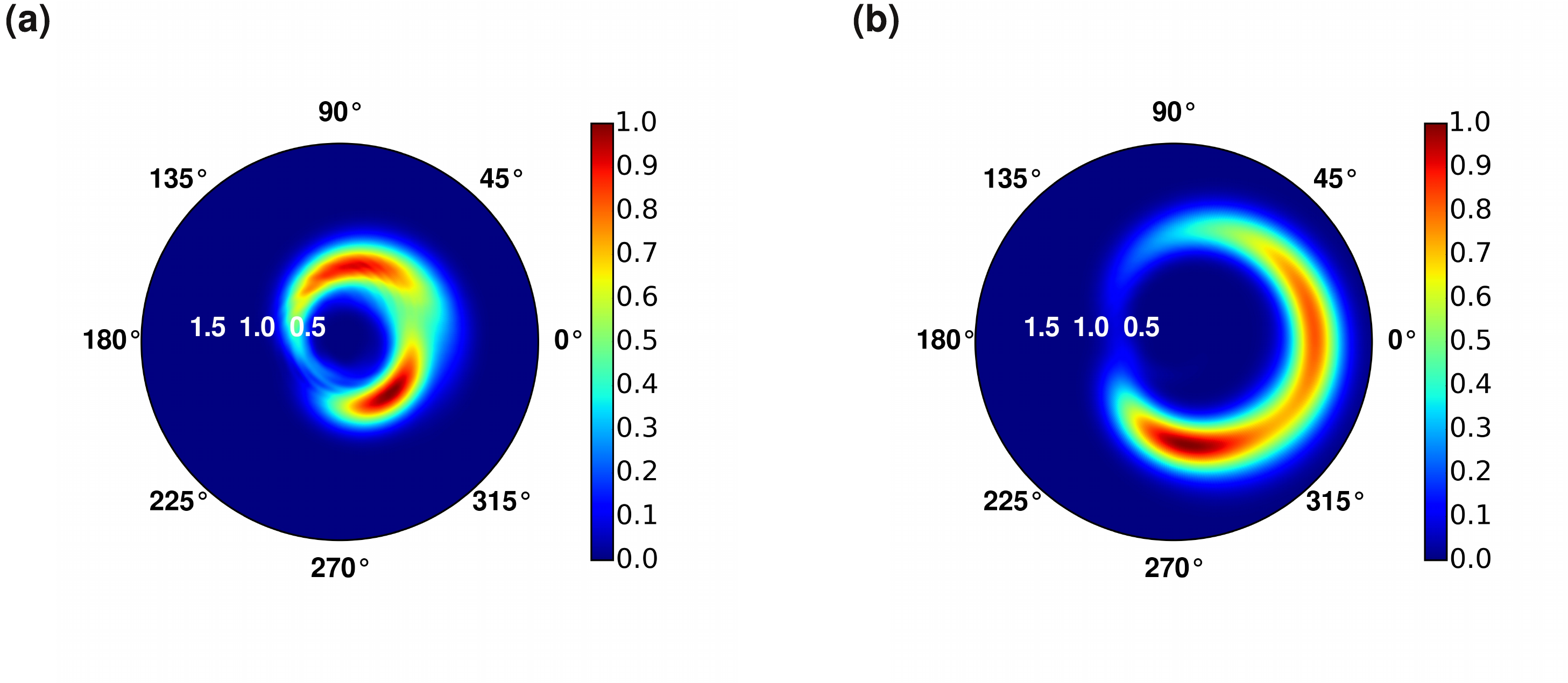}
\caption{ Angle- and energy- resolved spectra 
of two-electron ionization, for the same pulse shape 
as in the main text, for two-electron ionization of the model 4D two-electron system 
with $Z=+3$, and $I_p$=0.5 a.u. 
The calculations were done for $\lambda=800$nm,  $\epsilon=1$,
 and the same pulse shape as in the main text, using MCTDH method. 
The fields strengths are $F=0.04$ a.u. (a) and $F=0.0725$ a.u. (b)
} \label{FigSM5}
\end{figure}

{Very interesting are results for double ionization. The  two-electron spectra corresponding
to double ionization, are shown in   Fig.\ref{FigSM5}. At lower intensities, we
see that the ejection of the second electron is delayed relative to the first by about half a cycle.  At higher laser intensities,
the ejection angle between the two electrons is reduced. The two electrons depart with a delay of just about a quarter-cycle, consistent with the excitation of the second electron
during  the removal of the first\cite{Walters,keller3,arm2}.
Overall, we see correlation between the onset of positive additional deflection angle $\delta \theta$ in 
one-electron ionization and the one-set of double ionization.}

%TC:endignore
% \end{document}
>

\end{document}